\documentclass[preprint,12pt,authoryear]{elsarticle}

\usepackage{amssymb}
\usepackage{amsmath}

\journal{Communications in Nonlinear Science and Numerical Simulation}

\begin{document}

\begin{frontmatter}



\title{Fold catastrophe in breaking waves} 


\author{Francesco Fedele} 

\affiliation{organization={School of Civil and Environmental Engineering, Georgia Institute of Technology},
            addressline={790 Atlantic Drive}, 
            city={Atlanta},
            postcode={30332-0355}, 
            state={Georgia},
            country={USA}}

\begin{abstract}
We present a dynamical-systems perspective on wave breaking for ideal incompressible free-surface flows. By tracking the most energetic hotspot on the wave surface, we find that near breaking the surface slope $m$ evolves on a fast timescale governed by the small parameter $\epsilon = (\partial_z u)^{-1}$, the inverse vertical velocity gradient at the hotspot, while the focusing parameter $A = (U - C_e)/(U - C_{req})$ varies slowly and adiabatically. Here, $U$ is the horizontal fluid velocity at the energetic point, $C_e$ its propagation speed, and $C_{req}$ the equivalent crest speed. This slow–fast structure reveals a fold catastrophe in the $(m,A)$ space, whose boundary forms the geometric skeleton organizing the dynamics near breaking. Finite-time blowup occurs when the trajectory crosses this boundary, marking the onset of breaking.

The inception of breaking is further characterized by crossing the slope threshold $\theta_* = \arctan(\sqrt{2}-1) = 22.5^\circ$. This critical angle marks the maximum anisotropy that can be sustained between the Hessians of the velocity and pressure fields, reflecting an imbalance between kinetic and potential energy fluxes. The anisotropy of the velocity Hessian also gives rise to the classical $30^\circ$ slope observed at the inflection point of steep waves near breaking inception. The crest height is limited by the maximum excess of kinetic over potential energy that the flow can sustain, beyond which breaking becomes inevitable. Wave breaking can also be interpreted as a gravity analogue of a collapsing black hole, with apparent and event horizons representing the onset and inception of breaking.
\end{abstract}







\begin{keyword}
surface gravity waves, wave breaking, fold catastrophe.

\MSC[2020] 37E35 \sep \MSC[2020] 76B07 \MSC[2020] 76B15 \sep \MSC[2020] 58K35
\end{keyword}

\end{frontmatter}



\section{Introduction}
Breaking onset occurs when the fluid velocity at the crest just exceeds the crest speed, often causing fluid particles to detach subsequently from the surface~ \citep{phillips1966dynamics,PhillipsBanner1974,BannerPhillips1974}. When a wave breaks on a sloping beach, the fluid may plunge forward enclosing an air bubble or it may spill. In deep water, the spilling motion is typical, with a characteristic formation of a bulge and fluid elements sliding down the leading slope~ \citep{duncan1981Royalexperimental,Duncanqiao2001gentlebreaker_crestflow,duncan1999gentlebreaker_crest_profile}. At larger scales of breaking, air entrainment may take place and the breaking wave is clearly visible as a whitecap.

Recent studies have uncovered fundamental properties of wave breaking by examining the dynamics of wave groups by tracking specific moving points on their surface, such as crests and energetic hotspots. \cite{Banner2014CrestSlowdown}, through numerical simulations and field observations, showed that wave crests approaching their maximum height slow down significantly and either break at this reduced speed or accelerate forward unbroken. This crest slowdown is a generic feature that arises from the natural dispersion of waves~\citep{FedeleEPL2014,FedeleJFMcrestspeed2020}. It explains why breaking wave crests travel about $20\%$ slower than predicted by linear theory~\citep{RappMelville1990}. A following study by~\cite{BarthelemyJFM2018} identified a robust precursor to the breaking onset of wave crests. This is an inception point marking a state of no return beyond which the crest inevitably approaches the breaking onset and collapses. Through detailed numerical wave tank simulations, they found a robust threshold parameter for this inception valid for wave packets propagating either in deep or intermediate water depths. 
\cite{BarthelemyJFM2018} found that maximally tall non-breaking waves are clearly separated from marginally breaking waves when the parameter $B=U/C_r\approx 0.85$, where $U$ is the horizontal fluid velocity  and $C_r$ is the crest speed of the tallest wave~(see, also~\cite{SaketJFM2017} for an experimental validation). \cite{DerakhtiJGR2020} validated the proposed breaking threshold for surface gravity water waves in arbitrary water depth, including shallow water breaking over varying bathymetry. \cite{Touboul_Banner_2021} proved that it also holds for waves propagating in the presence of a constant background vorticity. \cite{Derakhti2018} also showed that the breaking strength of $2$D and $3$D gravity waves, measured by the energy dissipated per breaking event, is strongly correlated with the rate of change~$dB/dt$ at the breaking onset. Recent experimental work has examined the kinematic aspect of the breaking onset~\citep{Shemer2014LagrangianKinematics, Liberzon2017}, leading to novel detection methods~\citep{Liberzon_breaking2019} that identify breaking crest bulges by analyzing instantaneous frequency variations in the wave signal~\citep{Zimmermann_Seymour2002}. 

More recently, \cite{BoettgerJFM2023} identified a robust energetic signature of breaking inception by numerically solving for nonlinear wave packets under various forcing conditions, including cases with surface tension. They found an energetic~\emph{hotspot} on the forward face of the crest where the kinetic energy density attains its maximum~\citep{Boettger2024}. Crest breaking invariably occurred when the kinetic energy growth rate, measured at this hotspot, exceeded a critical threshold. This threshold, apparently dependent on the wave generation method, reflects an excessive convergence of kinetic energy at the crest that a wave cannot sustain. 
\cite{BoettgerJFM2023,BoettgerPRF2024} analyzed the geometric and kinematic properties of this energetic inception point finding a weak dependence on how waves are generated or forced. %

In contrast, \cite{McAllister2023} performed numerical simulations of $2$D inviscid, and irrotational water waves in the absence of surface tension. By tracking the surface point of maximum inflection, they found that the minimum local interface slope $m_{b} = \tan(\pi/6) \approx 0.5774$ at that point separates breaking from non-breaking waves. This threshold corresponds to an angle of $30^\circ$, matching the limiting steepness derived by~\cite{Stokes1847}. However, this threshold is not universal, as it strongly depends on how the waves are forced~\citep{duncan1983JFMbreaking}. Indeed, \cite{Boettger2025} noted that the critical slope~$m_b$ varies in the presence of surface tension and can reach angles approaching~$60^\circ$. They showed that surface tension effects are significant as the wave crest approaches its maximum steepness. For the steepest non-breaking waves, this manifests as a bulge on the crest tip that increases the local interface angle. 

A theoretical effort to explain the wave breaking phenomenon is related to the superharmonic instability of periodic Stokes waves~\citep{LonguetHigginsSuperHarm1978,Creedon_Deconinck_Trichtchenko_2022,DeconinckStokes2023, berti2024infinitely}. Recent studies have investigated the emergence of breaking water waves in deep water induced by perturbations to unstable Stokes waves~\citep{MansarBridges2025SHInstability,DeconinckStokes2023}, originally studied by~\cite{Cokelet1978}. In this regard, \cite{FedeleJFM2014} considered the dynamics of the compact Dyachenko-Zakharov~(cDZ) equation for unidirectional deep-water waves, a third-order model derived from the Euler equations~\citep{cDZ2011}. Perturbations to superharmonically unstable uniform wave trains of the cDZ equation in deep water evolve according to a modified Korteweg–de Vries equation, suggesting a possible primordial mechanism conducive to wave breaking, inherited from the Euler equations~\citep{FedeleJFM2014}.

Another complementary approach has been the study of breaking in wave systems governed by nonlinear dispersive equations. In this context, a wave breaks when its profile steepens until the slope becomes infinite in finite time, leading to the blow-up of smooth solutions. This behavior is of significant interest in fluid dynamics, particularly for the Euler equations modeling surface water waves~(see, e.g., \cite{Dyachenko2008}), as well as asymptotic models such as the \cite{Whitham1967}~equation among others~\citep{KalischWhitham2015}. \cite{Seliger1968} showed that if the initial wave profile is sufficiently asymmetric, it leads to breaking in the Whitham equation, with nonlinear steepening overcoming dispersion.  Seliger’s method focuses on tracking the maximum slopes at the inflection points on the forward and backward faces of a wave crest, where the curvature vanishes. He derived differential equations governing the evolution of the slope difference and showed this quantity becomes unbounded in finite time. \cite{Constantin1998} and \cite{Hur2017} later provided rigorous proofs of this result. Measuring a surface slope, or angle, to detect the breaking inception, or point of no return, of a wave is appealing. It is relatively simple compared to the $B$ threshold, which requires measuring both fluid and crest velocities. However, whether a universal slope threshold exists that separates non-breaking from breaking waves remains to be established.

In this work, we aim to provide further insight into the nature of wave breaking and the existence of thresholds for breaking inception. We adopt a dynamical-systems perspective and, from first principles, derive a slow–fast model that captures wave dynamics near breaking, as observed from the energetic hotspot~\citep{BoettgerJFM2023}. 

To derive this model, we first review the Euler equations for $2$D irrotational surface water waves. We then consider the evolution of local observables following a generic moving point on the surface. These observables evolve according to a system of coupled ordinary differential equations, which is generally not closed.  We then provide the physical insights that led to the identification of a relational fast timescale~$\tau$ near wave breaking~\citep{Rovelli1991TimeHypothesis,rovelli2004quantum,rovelli2018order}. This defines a \emph{steepening} time that is faster than the absolute Newtonian time~$t$ measured in the laboratory frame to observe waves. 
The wave evolution on the fast timescale~$\tau$, as observed at the energetic hotspot, is characterized by slow variations of kinematic variables and rapid changes of geometric parameters such as the surface slope. A slow–fast model naturally arises from first principles, revealing breaking as a fold catastrophe~\citep{Thom1975,zeeman1976catastrophe,arnold1984catastrophe,Gilmore1993}. 

Finally, we discuss our results in relation to the numerical findings of~\cite{BoettgerJFM2023} and~\cite{McAllister2023}, and outline how wave breaking can be viewed as the gravity analogue of collapsing black holes.

\section{Governing Equations}

Consider an incompressible and irrotational inviscid fluid~(water) of density~$\rho$ sharing a zero-stress free surface boundary with another fluid~(air) of vanishing density and viscosity. Waves propagate over a finite depth~$d$. The Euler equations govern the evolution of the velocity field 
$\mathbf{u} = (u,w)$ beneath the free surface $z = \eta(x,t)$ of unidirectional waves in water of depth~$d$. We adopt a Cartesian coordinate system with upward vertical axis~$z$ and horizontal axis~$x$ pointing from left to right. Partial differentiation is denoted interchangeably by~$\frac{\partial f}{\partial x}=\partial_x f = f_x$.

The kinematic boundary condition
\begin{equation}\label{kin}
   \eta_t + u\, \eta_x = w\,, \quad z = \eta\,,
\end{equation}
imposes that fluid particles do not penetrate above the free surface, but remain on it.
The dynamic condition expresses momentum conservation throughout the fluid domain:
\begin{equation}\label{Euler}
D_t u = -\partial_x \left(p/\rho\right)\,, \qquad  
D_t w = -\partial_z \left(p/\rho + g z\right)\,,
\end{equation}
where the Lagrangian derivative is $D_t = \partial_t + \mathbf{u} \cdot \nabla$, with $\nabla = (\partial_x, \partial_z)$, $p$ is the pressure field and $g$ is the gravitational acceleration.

The continuity of normal stresses across the surface imposes a dynamic boundary condition: the fluid pressure~$p$ along that boundary is equal to the air pressure~$p_a$, which is a constant, 
\begin{equation}\label{Pdyn}
    p(x,z,t)|_{z=\eta} = p_a\,,
\end{equation}
For varying bathymetry~$z_b=h(x)$, the normal fluid velocity at the sea floor must vanish, i.e.,  $\mathbf{u} \cdot \mathbf{n}_b=0$, where the unit normal~$\mathbf{n}_b=(-h',1)/\sqrt{1+h'^2}$ and $h'=dh/dx$ denotes differentiation with respect to~$x$. 

Incompressibility imposes
\begin{equation}\label{incompressible}
    \nabla \cdot \mathbf{u} =\partial_x u + \partial_z w = 0\,,
\end{equation}
and irrotationality requires zero vorticity
\begin{equation}\label{irrotational}
   \partial_x w = \partial_z u\,,
\end{equation}
everywhere in the fluid region.  This implies that the Eulerian velocities are harmonic functions, i.e., $\Delta u=\Delta w =0$, where the Laplacian $\Delta =\nabla^2 = \partial_{xx} + \partial_{zz}$.  Equivalently, one may introduce a velocity potential $\phi(x,z,t)$ such that $\mathbf{u} = \nabla \phi$, in which case the Euler equations can be rewritten in terms of $\phi$ and $\eta$.

Hereafter, we consider inviscid and irrotational waves in deep waters with no surface tension. To make physical variables dimensionless, we use the rescaling
\begin{equation}\label{resclalingdeepwater}
 x\rightarrow \frac{x}{L},\quad z\rightarrow \frac{z}{L}, \quad t\rightarrow \frac{t}{T}, \quad \mathbf{u}\rightarrow \frac{\mathbf{u}}{c},\quad \eta\rightarrow \frac{\eta}{L},\quad p\rightarrow\frac{p}{\rho c^2}\,,\quad g\rightarrow \frac{gT}{c}\,,
\end{equation}
where $L,T$ are a characteristic wavelenghth and wave period,  and the speed $c=L/T$. The Euler equations remain invariant under this rescaling, except that the gravitational acceleration transforms as $g \to gT/c$. Following~\cite{BarthelemyJFM2018}, we choose $L = 2\pi/k$ and $T = 2\pi/\omega$, where the wave frequency~$\omega$ and wavenumber~$k$ satisfy the linear dispersion relation $k = \omega^{2}/g$ in deep water. Then, $c = \omega/k=\sqrt{gk}$ is the phase speed and $g = 2\pi$.

\subsection{Kinetic energy transport}
In this study, we will need the Bernoulli equation and the kinetic energy transport equation. From Eq.~\eqref{Euler}, energy conservation leads to the Bernoulli equation at the free surface:
\begin{equation}\label{Bernoulli}
\phi_t + g\, \eta + k_e = 0\,, \qquad z = \eta\,,
\end{equation}
where $k_e =|\mathbf{u}|^2 / 2=(u^2+w^2)/2$ is the kinetic energy density per unit mass.  
The kinetic energy transport equation~\citep{Tulin2007,FedeleJFM2016} 
\begin{equation}
\partial_t k_e + \nabla \cdot \mathbf{F}_{k_e}  = 0\,,
\end{equation}
where the kinetic energy vector flux~$\mathbf{F}_{e} =-\phi_{t} \mathbf{u}$, and the Eulerian kinetic energy flux speed follows as
\begin{equation}\label{Kefluxspeed1}
\mathbf{C}_{k_e} =\frac{\mathbf{F}_{k_e}}{k_e} = -\frac{ \phi_{t}}{k_e} \mathbf{u}\,.
\end{equation}

At the free surface,
\begin{equation}\label{Kefluxspeed}
\mathbf{F}_{k_e} = \mathbf{u} \left( g \eta + k_e \right)\,,\qquad\mathbf{C}_{k_e}=\mathbf{u}\left(1+\frac{g \eta}{k_e}\right)\,, \qquad z = \eta\,,
\end{equation}
where we have used Bernoulli~\eqref{Bernoulli}. 

\subsection{Pressure relations}
Since the pressure is uniform on the free surface, its tangential derivatives of any order vanish there. The first derivative yields the relation
\begin{equation}\label{Pressurerelation1}
  \nabla p \cdot\mathbf{t}\sim R=p_x + \eta_x \,p_z = 0\,,\qquad z=\eta\,,  
\end{equation}
and thus $\partial_x^n R=0$ at $z=\eta$, for all $n>1$. For example, for $n=2$ one obtains 
\begin{equation}\label{Pressurerelation2}
\nabla (\nabla p\cdot \mathbf{t}) \cdot\mathbf{t}\sim \partial_x^2 R= p_{xx} + 2 \eta_x \,p_{xz} +\eta_x^2 \,p_{zz}  + \eta_{xx}\, p_z = 0\,,\qquad z=\eta\,,
\end{equation}
where the unit tangent~$\mathbf{t}=(1,\eta_x)/\sqrt{1+\eta_x^2}$. 

Taking the divergence of the momentum equations~\eqref{Euler} yields the Poisson equation for the pressure
\begin{equation}\label{eq:PoissonP}
\Delta p = -\rho (u_x^2 + 2u_z w_x + w_z^2)\,. 
\end{equation}
Using irrotationality and incompressibility ($w_x=u_z,\,w_z=-u_x$), this becomes
\begin{equation}\label{eq:PoissonP2}
\Delta p = -2\rho (u_x^2 + u_z^2). 
\end{equation}

\subsection*{Parity-reflection symmetry}
The inviscid, irrotational Euler equations with a flat bottom, or in deep water are invariant under the parity transformation~$x \mapsto -x$, 
\begin{equation*}
\,\,\eta(x,:) \mapsto \eta(-x,:)\,,
\,\, u(x,:) \mapsto -u(-x,:),
\,\, w(x,:) \mapsto w(-x,:),
\,\, p(x,:) \mapsto p(-x,:),
\end{equation*}
That is, if $(\eta,u,w,p)$ is a solution, then so is its parity-reflected counterpart. The horizontal velocity $u$ also changes sign because it lies along the reflected axis, while the signs of vertical velocity $w$, the pressure $p$, and the free-surface elevation $\eta$ remain unchanged.
Physically, it means there is no preferred direction along the $x$-axis for wave propagation. A right-traveling wave  and its mirror image, a left-traveling wave, are both solutions of the Euler equations. 

The symmetry is broken by physical effects that distinguishes left from right propagation, such as a background shear current, rotation (Coriolis force), or sloping and irregular seabed bathymetry. For example, a steady current breaks symmetry, because right-going and left-going waves see different Doppler shifts. Wind forcing and dissipation, surface tension gradients add directionality and also break the symmetry.

\subsection{Vorticity at the Free Surface}
For an inviscid and irrotational fluid~(water), vorticity in the bulk vanishes identically by Kelvin's circulation theorem, and Eq.~\eqref{irrotational} holds everywhere in the interior of the fluid. However, the zero-stress condition at the surface allows fluid parcels to rotate with the evolving unit normal vector~\citep{LonguetHiggins1998}, thereby generating surface vorticity~\citep{FedeleJFM2016}:
\[
\omega= -2\,\frac{\eta_{xt} + u \eta_{xx}}{1+\eta_x^2}\,.
\] 

The free surface also supports a \emph{vortex sheet} with strength 
\(\gamma = \mathbf{u}_t^+ - \mathbf{u}_t^-\), defined as the jump in tangential velocity across the interface~\citep{Terrington2020,Brons2014,lundgren1999generation}. Here, $\mathbf{u}_t^+$ and $\mathbf{u}_t^-$ denote the tangential velocity just above and just below the surface, respectively. 
Because the density and viscosity of the upper fluid (air) are effectively negligible, the sheet strength reduces to~$\gamma = - \mathbf{u}_t^-$, that is, minus the tangential velocity of the water at the surface. This concentrated vortex layer formally contains all the vorticity generated by surface curvature and by the unsteady deformation of the free boundary. 

In a viscous fluid, surface-generated vorticity diffuses into the interior, producing shear layers~\citep{phillips1966dynamics,fedele2022momentary}. These can grow and interact with the vortex sheet while conserving the total circulation induced both by the vortex sheet and by the viscous interior flow~\citep{Terrington2020,Brons2014,lundgren1999generation}.  

In the present study, by contrast, the flow is taken to be potential. In this case, surface vorticity remains confined to the free surface and does not diffuse into the bulk, since viscous effects are absent. Therefore, the vortex sheet and surface vorticity are dynamically irrelevant for the evolution of the free surface and the underlying irrotational flow. Consequently, when approaching the surface from below, the irrotationality condition in Eq.~\eqref{irrotational}, i.e., $\partial_z u = \partial_x w$, also holds at the free surface.

\section{Tracking observables on the free surface to probe breaking}

We study the surface dynamics by tracking the values of physical fields, or \emph{observables}, in time at a specific moving point $Q$ on the wave surface.
This defines a trajectory that is not Lagrangian, as it follows a geometrically distinguished point on the wave profile rather than a fluid particle. For example, \cite{BarthelemyJFM2018} investigated breaking inception by tracking a wave crest. \cite{Seliger1968} and \cite{McAllister2023} tracked surface points of maximum inflection, and \cite{BoettgerJFM2023} followed surface points of maximum kinetic energy.

The relevant local observables tracked at $Q$ we consider are the surface slope $m=\eta_x|_Q$, the local fluid velocities~$(U,W)=(u|_Q,w|_Q)$, the gradient components
\[
(U_x,U_z,U_{xx},U_{xz},U_{zz})=(u_x,u_z,u_{xx},u_{xz},u_{zz})\big|_Q\\,
\]
and the surface curvature proportional to~$\eta_{xx}|_Q$. We also consider the following space derivatives of the pressure field
\[
(P_x,P_z, P_{xx},P_{xz},P_{zz})=(p_x,p_z,p_{xx},p_{xz},p_{zz})\big|_Q\,.
\]
The kinetic energy density per unit mass tracked at $Q$ is given in terms of the above observables as~$K_e=(U^2+W^2)/2$. 

The time evolution of these quantities leads to a system of ordinary differential equations, which is not closed~(see~\ref{appAODE}). The time evolution of each of these local observables depend on the others due to the nonlinear nature of the Euler equations.
Completing the ODE system would require adding dynamical equations for all these related observables, which is an extremely cumbersome task. A practical solution would involve a truncation procedure to achieve closure.
Nevertheless, near the breaking regime, a slow-fast timescale separation occurs and the time evolution of a few key observables is sufficient to characterize wave breaking as a fold catastrophe bifurcation. 


\subsection{Dimensionless wave parameters}
We define the following dimensionless parameters tracked at a moving point~$Q$ on the free surface to characterize the wave dynamics of steep waves and near breaking:
\begin{equation}\label{s&m2}
s = \frac{W}{U}\,,\qquad m_2 =\frac{U_x}{U_z}\,.
\end{equation}
The ratio~$s$ is kinematic and measures the relative importance of vertical fluid velocities with respect to the horizontal fluid flow. The slope of the fluid velocity vector~$\mathbf{u}$ is given by $\tan \theta_{\mathbf{u}}=s$, where $\theta_{\mathbf{u}}$ is the angle  with respect to the horizontal.  The ratio~$m_2$ is a momentum parameter
\begin{equation}\label{m2_meaning}
    m_2=\frac{U\,U_x}{U\,W_x} \sim \frac{\text{horizontal advection of horizontal momentum}}{\text{vertical transport of horizontal momentum}}\,.
\end{equation}
which quantifies the relative contribution of horizontal advection to vertical transport of horizontal momentum density~$\rho u$ towards the surface, and $W_x = U_z$ because the fluid is assumed irrotational. Note that the slope of the velocity gradient~$\nabla U$ is $\tan\theta_{\nabla U}=-1/m_2$, where~$\theta_{\nabla U}$ is the angle with respect to the horizontal. 

We also consider the focusing parameter
\begin{equation}\label{A}
A=    \frac{(U - C)\, \eta_{xx}|_Q}{U_z} = \frac{U - C}{U - C_{req}} \,,
\end{equation}
with $C_{req}$ being the equivalent crest speed at $Q$, i.e., the speed as if there were a crest at $Q$, and derived in~\ref{appC_eqcrestspeed}. In general, near a focusing wave crest, $C_{req} \geq C$ at $Q$, and when $Q$ passes through the wave crest, $C = C_{req}$ and $A = 1$ where the maximum crest height is attained at focus.\footnote{From the definition of $A$, $C = (1 - A) U + A C_{req}$, meaning $C$ is a weighted average between the fluid speed $U$ and the equivalent crest speed $C_{req}$.} In the range of $0<A<1$, we have $U < C < C_r$ and $U/C<1$ and $C_{req} / C > 1$. At the breaking onset, where $U\rightarrow C$, $A\rightarrow 1$. 

Small $m_2$ implies that vertical transport dominates and momentum is efficiently transferred upward toward the crest. 
Near a very steep wave crest, and especially approaching breaking, vertical shear $U_z$ tends to be large and horizontal gradients $U_x$ and vertical fluid velocities~$W$ are smaller. Hence $m_2$ and $s$ are typically small, indicating that vertical momentum transport dominates. Physically, this is linked to the formation of jets or overturning after the breaking onset.

\subsection{Intrinsic timescale}\label{Intrinsictimescale}
We rely on the following simple physical picture for the dynamics of waves near breaking, referring to non-breaking steep-to-maximally-steep waves and breaking waves.
Imagine tracking a special moving point~$Q$ on the wave surface, such as the energetic hotspot~\citep{BoettgerJFM2023}, and the associated speed $C=C_e$. Moving with that point, the fluid beneath the surface appears to flow more slowly than observed from a fixed frame, with a small relative velocity~$U-C_e$. This motivates the definition of the small parameter 
\begin{equation}\label{beta}
    \beta = C_e/U-1 = 1/B - 1\, \ll 1\,, \text{ near breaking},
\end{equation}
where $B = U/C_e$. Near the breaking onset of a wave crest, $B \rightarrow 1$, corresponding to $\beta \rightarrow 0$ from above. 

\begin{figure}[htbp]
  \centering
  \includegraphics[width=0.85\linewidth]{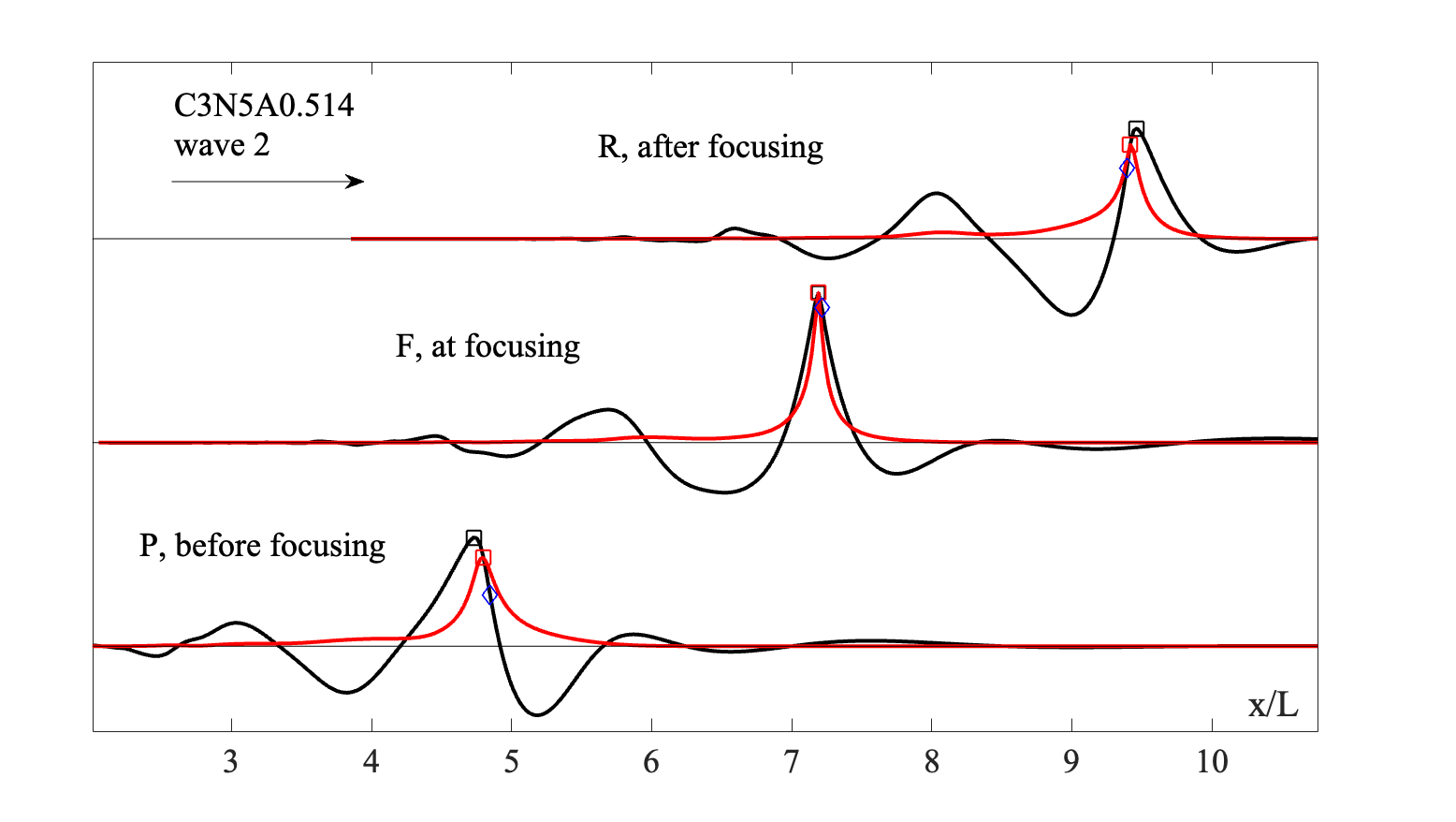}\\[1ex] 
  \centering
    \hspace*{-0.5cm}\includegraphics[width=1.1\linewidth]{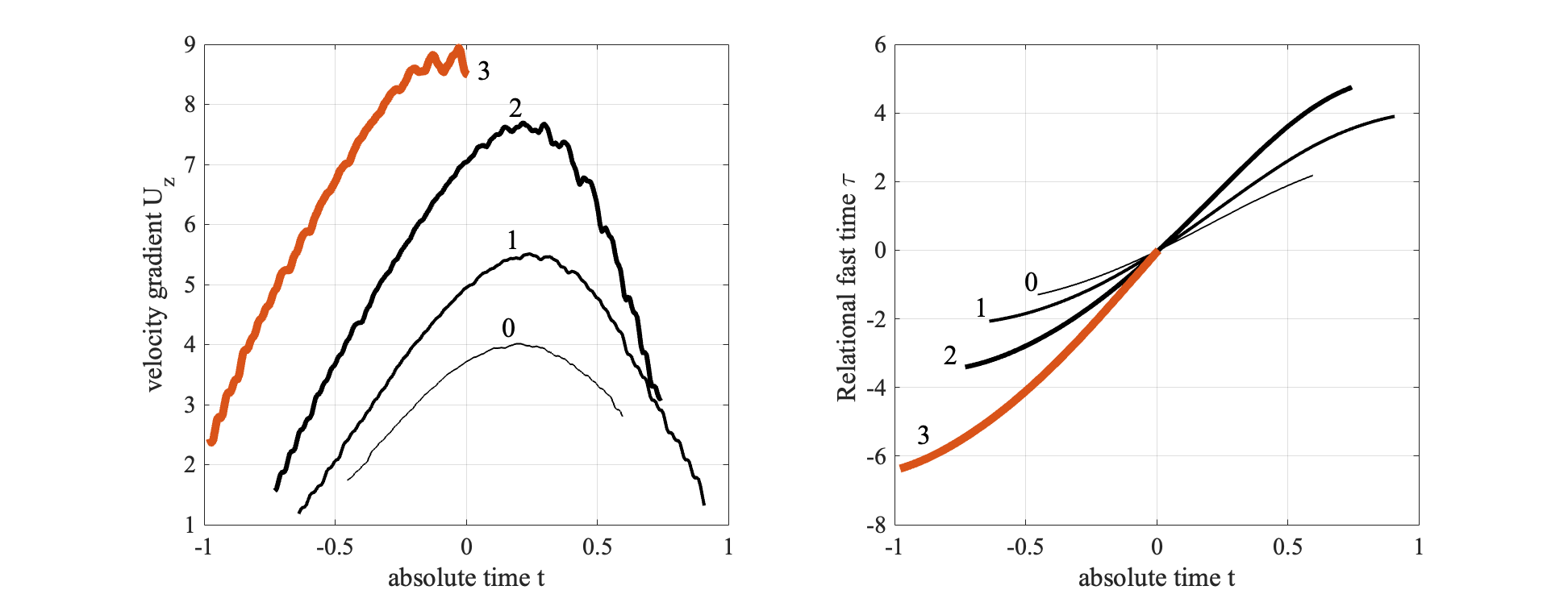}\\[1ex]
    
  \caption{Top panel: snapshots of the (black) surface elevation~$\eta(x,t)$ and (red) kinetic energy~$\tilde{k}_e(x,t)$ of the largest non-breaking wave (labeled~$2$) in the simulated \emph{right-going} wave group $C3N5A0.514$~\citep{BarthelemyJFM2018} at the pre-focusing~($P$), focusing~($F$), and post-focusing~($R$) instants, in the middle top panel of Fig.~\ref{FIG3fold}. The energetic hotspot~(red square) is located at the front of the wave before focusing and moves to the back after passing through the crest~(blue square). The inflection point is also depicted~(blue diamond). Bottom panel: (left) time profiles of the horizontal velocity gradient~$U_z$ tracked at the energetic hotspot of the four major successive waves of the simulated \emph{right-going} group $C3N5A0.514$~\citep{BarthelemyJFM2018}. These are labeled as $0,1,2,3$ with varying thickness to denote the increasing amplitude, and wave~$3$ is breaking; (right) relational fast time~$\tau$ as function of the absolute time~$t$.}\label{FIG2waveprofiles+crestgroupspeeds}
\end{figure}

The velocity gradient $U_z = \partial_z u |_Q$ measured at the moving point~$Q$ governs the rate of change of the dynamics near breaking. As a focusing wave steepens, $U_z$ increases in magnitude, indicating that vertical motion transports significant horizontal momentum toward the surface. This process amplifies the horizontal fluid motion and enhances the convergence of kinetic energy near the crest during wave steepening. Such a concentration of horizontal momentum and energy flux ultimately leads the wave to break~\citep{BoettgerJFM2023,BoettgerPRF2024}. This trend is observed in numerical simulations of unsteady breaking and non-breaking wave groups~\citep{BarthelemyJFM2018}~(see bottom-left panel of Fig.~\ref{FIG2waveprofiles+crestgroupspeeds}). In the simulations, the velocity gradient $U_z$ increases as the waves steepen, reaching a maximum at breaking, where the numerical solver stopped because it was not designed to follow the post-breaking phase.
The relavance of velocity gradients to wave breaking was first emphasized by~\cite{PhillipsBanner1974}, who showed that thin shear layers induced by wind can trigger breaking at reduced amplitudes~\citep{BannerPhillips1974}. 

Following~\cite{Rovelli1991TimeHypothesis,rovelli2004quantum,rovelli2018order}, we define the dimensionless relational fast time~$\tau$\footnote{Time, in Rovelli’s relational framework, measures change, not an absolute flow. Temporal order arises only through correlations between observables. The surface elevation of a wave changes with respect to the reading of some other observable~(clock), which in our case we choose as related to the vertical velocity gradient.} that measures the relevant changes of the dynamics in relation to \emph{steepening} as observed at a moving point~$Q$ on the surface:
\begin{equation}\label{timescaletau}
    \frac{d\tau}{dt} = U_z\,, \qquad \tau(t)=\int^t_{t_0} U_z(t')\, \text{d}\,t'\,,
\end{equation}
where $t_0$ is a convenient initial time after which $U_z \ge 0$, ensuring that $\tau$ increases monotonically~(see bottom-right panel of Fig.~\ref{FIG2waveprofiles+crestgroupspeeds}, note the smoothness of $\tau(t)$ as a result of the integration). This definition applies to waves propagating from left to right along the $x$-axis, where $U_z > 0$ near the crest of steep waves.\footnote{For waves propagating in the opposite direction (right to left), $U_z < 0$ and one sets $\frac{d\tau}{dt} = -U_z$ to maintain a monotonically increasing $\tau$.} As waves steepen, the horizontal velocity gradient $U_z$ measured at $Q$ grows, eventually becoming very large and formally diverging at the onset of breaking, for instance during the formation of jets associated with spilling.  In other words, the internal clock associated with the crest-steepening dynamics ticks more quickly than the external time we use to observe the waves.
This motivates defining near breaking, the small parameter
\begin{equation}\label{epsilon}
    \epsilon = \frac{1}{U_z}\,, \qquad 0 < \epsilon \ll 1\,, \text{ near breaking},
\end{equation}
which separates slow and fast dynamics on the $\tau$-timescale as waves steepen or approach breaking. 

Both parameters $\beta,\epsilon$ are small near breaking, and~$\beta\sim O(\epsilon)$, estimated from numerical simulations~\citep{BarthelemyJFM2018}. 
For very steep wave crests near the breaking inception, where $U_z\gg 1$~($\epsilon\ll 1$), our analysis indicates that the focusing parameter $A$ in Eq.~\eqref{A} is bounded and of order~$O(1)$. 
Near breaking, this implies that $\eta_{xx}$, proportional to surface curvature, grows as fast as $O(\epsilon^{-2})$ to keep $A$ finite. Here, the notation $O(\epsilon^n)$ denotes terms of order $\epsilon^n$ or higher.

The physical picture is as follows: as waves steepen, the kinematic and dynamical fluid variables observed from the moving point~$Q$ evolve slowly on the $\tau$-timescale, in contrast to the rapid variations of geometric quantities such as surface slope and curvature. 

We demonstrate that this timescale separation occurs when the special moving point~$Q$ is chosen to coincide with the energetic hotspot~\citep{BoettgerJFM2023,BoettgerPRF2024}~(for details, see~\ref{appAslowfastODE}). It captures the essential dynamics near the onset of breaking and it reveals a robust fold geometry that guides the evolution of steep waves near breaking. 

Our analysis is grounded in first principles and supported by numerical simulations of nonlinear wave groups by~\cite{BarthelemyJFM2018}. 
The same separation could not be conclusively established when following other surface points, such as the inflection point~\citep{McAllister2023}.

\section{Following the energetic hotspot: the slow-fast model}

Consider the $1$D kinetic energy density field
\begin{equation}\label{1dkefield2}
    \tilde{k}_e(x,t)=k_e(x,z=\eta(x,t),t)\,,
\end{equation}
evaluated at the free surface $z=\eta(x,t)$, where $k_e=(u^2+w^2)/2$ is the kinetic energy density of the fluid~(see~\ref{appAKe}).  Drawing on~\cite{BoettgerJFM2023}, we track the most energetic point~$Q$ on the wave surface, defined as the location where $\tilde{k}_e$
attains its maximum, that is where the spatial kinetic energy gradient~$\partial_x\tilde{k}_e=0$. This energetic point can be regarded as the \emph{crest} of the kinetic energy \emph{surface }~$\tilde{k}_e(x,t)$.  The speed~$C_e$ of the energetic hotspot reduces to the group velocity for linear wave packets. 

To highlight the relating motion of the wave surface crest and energetic point, or hotspot, we consider the simulated breaking wave group~$C3N5A0.514$ reported in~\cite{BarthelemyJFM2018}. The four major successive waves of the group are labeled in order of increasing amplitude, with wave~$3$ undergoing breaking. The evolution of the second-largest wave (wave~$2$) and its kinetic energy surface is shown in the top panel of Fig.~\ref{FIG2waveprofiles+crestgroupspeeds}. As the wave grows in amplitude (point~$P$, pre-focusing phase), the energy crest is located on the front face of the wave, slightly behind the surface crest. At focus (point~$F$), where the wave attains its maximum height, the surface crest and the energetic point coincide at the same location. After focus (point~$R$), the energetic point shifts to the back face of the wave, behind the surface crest.  The other waves in the group have similar evolution, except wave~$3$ that breaks near the focusing point~(see \citep{BarthelemyJFM2018} for more details). 

During the growth phase, the surface crest slows down~\citep{FedeleJFMcrestspeed2020, Banner2014CrestSlowdown,FedeleEPL2014}, while the energetic point speeds up. However, the acceleration of the energetic point is insufficient to prevent the surface crest from overtaking it. After focus, the surface crest accelerates, whereas the energetic hotspot decelerates. The corresponding crest and energetic point speeds are shown in the bottom-right panel of Fig.~\ref{FIG2waveprofiles+crestgroupspeeds}. 

\subsection{The slow-fast model}
The energy gradient $\partial_x \tilde{k}_e$ vanishes at the energetic hotspot~$Q$. Thus, Eq.~\eqref{KEspacegradient} in~\ref{appAKe}) yields
\begin{equation}\label{m2eq}
m_{2}|_Q = \frac{U_x}{U_z} = \frac{U m + W}{W m - U} = \frac{m + s}{m s - 1}\,,
\end{equation}
where $m_2$ and $s$ are the momentum and kinematic parameters defined in Eq.~\eqref{s&m2}, and $m$ is the surface slope at~$Q$. Tracking the energetic hotspot determines the momentum transport through $m_2$ as a function of the kinematic and geometric observables $m$ and $s$. 

In~\ref{appAslowfastODE}, we proved that a slow-fast regime occurs near breaking around the maximum  $A=1+O(\epsilon)$, and along the path~$\Gamma_2$ in the $(m,s)$ plane where $m_2\sim(m+s)=0$~(see Eq.~\eqref{gamma_2}). This is where the vertical transport of horizontal momentum density dominates the horizontal advection during focusing when a wave reaches its maximum height, or breaks~(see Eq.~\eqref{m2_meaning}). Numerical simulations confirm this trend, as illustrated in Figure~\ref{ms_domain}, and is further discussed in section~\ref{Sec:Point of no returm}.  

A detailed balance near the breaking onset is found when
\begin{equation}\label{scalinga}
m,s\sim O(\epsilon^{1/2})\,,\quad A-1\sim O(\epsilon)\,,\quad C_e-C_{req}\sim O(\epsilon^{3/2})\,,\quad  P_z,U\sim~O(1)\,,
\end{equation}
and second order pressure derivatives scaling as
\begin{equation}\label{scaling2a}
P_{xx},P_{zz}~\sim~O(\epsilon^{-2}),\qquad P_{xz}\sim~O(\epsilon^{-3/2})\,,
\end{equation}
to balance the pressure equation~\eqref{eq:PoissonP} and pressure relation~\eqref{Pressurerelation2}. 
From Eq.~\eqref{Ce2}, near the breaking onset the hotspot velocity $C_e=C_{req}+O(\epsilon^{3/2})$ if $P_{xx}=P_{zz}+O(\epsilon^{3/2})$ and
\begin{equation}\label{scaling3a}
U_{xx},U_{zz},U_{xz},P_{xz}\sim~O(\epsilon^{-3/2})\,.
\end{equation}
Then, $dC_e/d\tau=dC_r/d\tau\sim O(\epsilon^{3/2})$. Using these scalings, in~\ref{appB_hotspot} a closed slow-fast system is derived in  the variables $(m,A)$~(see Eq.~\eqref{ODEdim4}), 
\begin{equation}\label{ODEdim4b}
\left\{
\begin{aligned}
&\frac{dm}{d\tau} = F(m,A)=-m^2 + 1 - A\,, \\[6pt]
&\frac{dA}{d\tau} = O(\epsilon^{3/2})\,,\\
\end{aligned}
\right.
\end{equation}
where $0<\epsilon\ll 1$. Here, $m$ is a fast variable and $A$ adiabatically slow. $R=m+s\sim O(\epsilon)$ is also slow variable and the $O(1)$~pressure gradient
\begin{equation}
    P_z/\rho= U U_z \left(1- A+\beta-m^2\right)-g\,.
\end{equation}
Here, $\beta= -A/(\epsilon\, \eta_{xx})\sim O(\epsilon)$ because $\eta_{xx}\sim O(\epsilon^{-2})$. 
And, $\beta$ is also a slow variable with $d\beta/d\tau\sim O(\epsilon^{3/2})$ along with the hotspot speed $C_e$ and equivalent crest speed $C_{req}$. Note that the limiting value of $P_z$ as $\epsilon\rightarrow 0$ is a special limit of the form $0\cdot\infty$. The region of validity of the fast-slow model is near the breaking region of the $m,A$ plane where $A-1\sim O(\epsilon)$ and $m\sim O(\epsilon^{1/2})$. In practice, the model captures the wave dynamics near breaking fairly even beyond its formal range of validity, as will be shown below. 
The hotspot condition of maximum kinetic energy decouples the evolution of observables in a distinctive way: the slope evolution does not depend explicitly on pressure forces, but only on the wave parameter~$A$.

\begin{figure}[htbp]
  \centering
 \hspace*{-2.5cm} \includegraphics[width=1.36\linewidth]{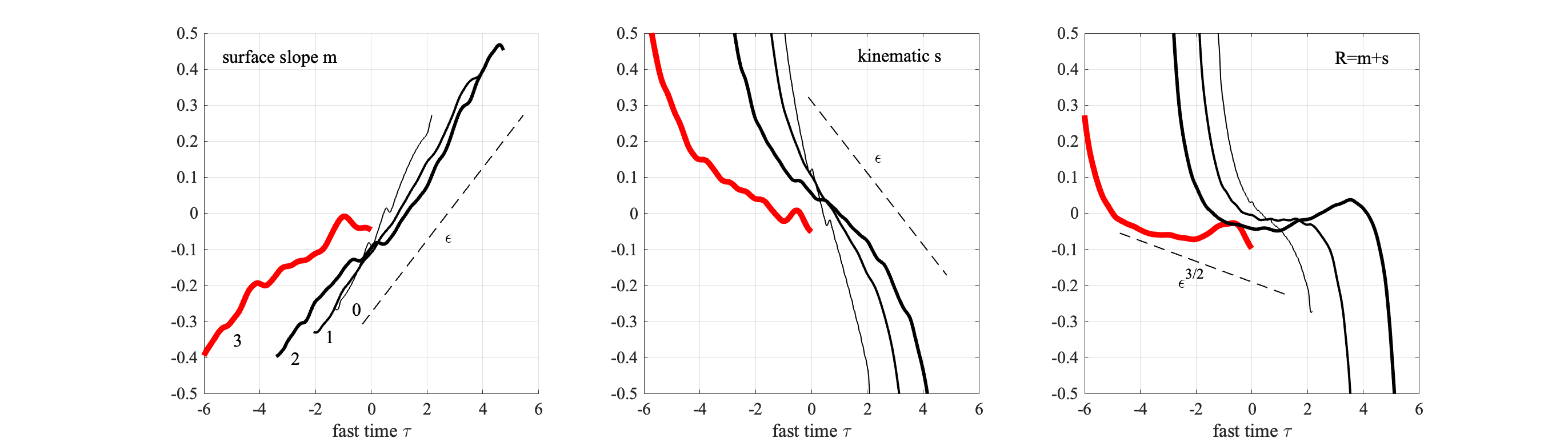}\\[1ex]
  \centering
  \hspace*{0.2cm}\includegraphics[width=0.73\linewidth]{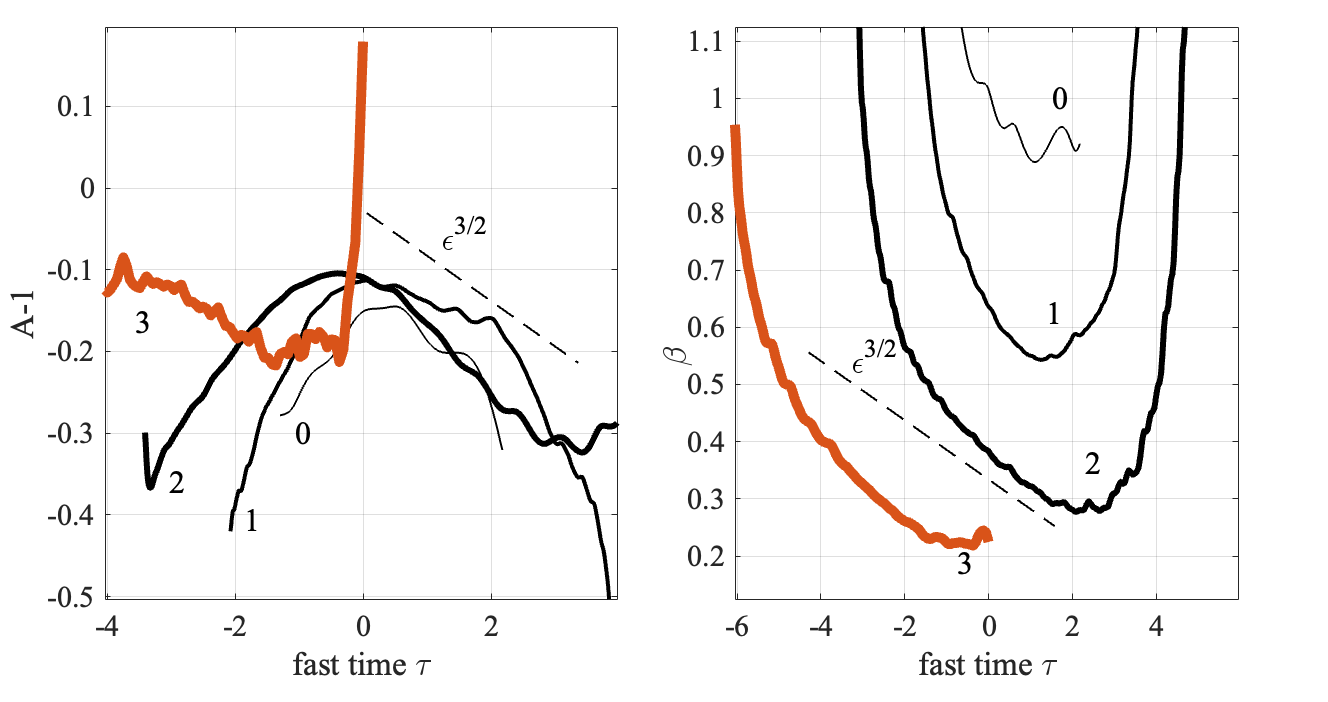}\\[1ex] 
  
\caption{Top panels: time profiles of the wave parameters~$m$, $s$ and $R=m+s$ tracked at the energetic hotspot of the four major successive waves of the simulated \emph{right-going} group $C3N5A0.514$~\citep{BarthelemyJFM2018}. Bottom panels: same for the slow parameters~$A$ and $\beta=C_e/U-1$ tracked at the energetic hotspot.  Waves are labeled as $0,1,2,3$ with varying thickness to denote the increasing amplitude, and with wave~$3$ breaking. Characteristic slopes are also shown (red lines) and quantified in terms of $\epsilon=1/U_z\approx 0.1$.}
  \label{FIG2_fastslow}
\end{figure}

Such slow-fast timescale separation is observed in numerical simulations of nonlinear wave groups~\citep{BarthelemyJFM2018,Boettger2024}. The top panels of Fig.~\ref{FIG2_fastslow} depict the time profiles of the wave parameters~$m$, $s$ and $R=m+s$ tracked at the energetic hotspot of the four major successive waves of the simulated \emph{right-going} group $C3N5A0.514$~\citep{BarthelemyJFM2018}. These are labeled in order of increasing amplitude, with wave~$3$ breaking.  Near breaking,  $R=m+s$ varies adiabatically,  more slowly than the surface slope~$m$ and the kinematic parameter $s$ over the fast time~$\tau$. Similarly, the focusing parameter $A$ and $\beta=C_e/U-1$ are also slow variables~(see  bottom panels of Fig.~\ref{FIG2_fastslow}.  Thus, the slow variable~$A$ acts as an adiabatic forcing for the fast geometric variable~$m$. 

Introducing the slow time scale $\tau_2=\epsilon^{1/2}\tau$ suggests a multiple scale perturbation approach for extending the region of validity of the fast-slow model beyond the breaking region. We conjecture that as the wave approaches the breaking onset, its effective dynamics slow down, analogous to approaching a collapsing black hole~(see section~$\S$~\ref{sec:BH}). Thus, higher-order pressure-gradient terms are fixed by requiring the next-order slowness of these variables in the hierarchical perturbation expansion, without the need to solve for the pressure equation~\eqref{eq:PoissonP} that couples pressure and velocity fields non-locally. 

\begin{figure}[htbp]
  \centering
  \hspace*{-0.5cm} 
  \includegraphics[width=1\linewidth]{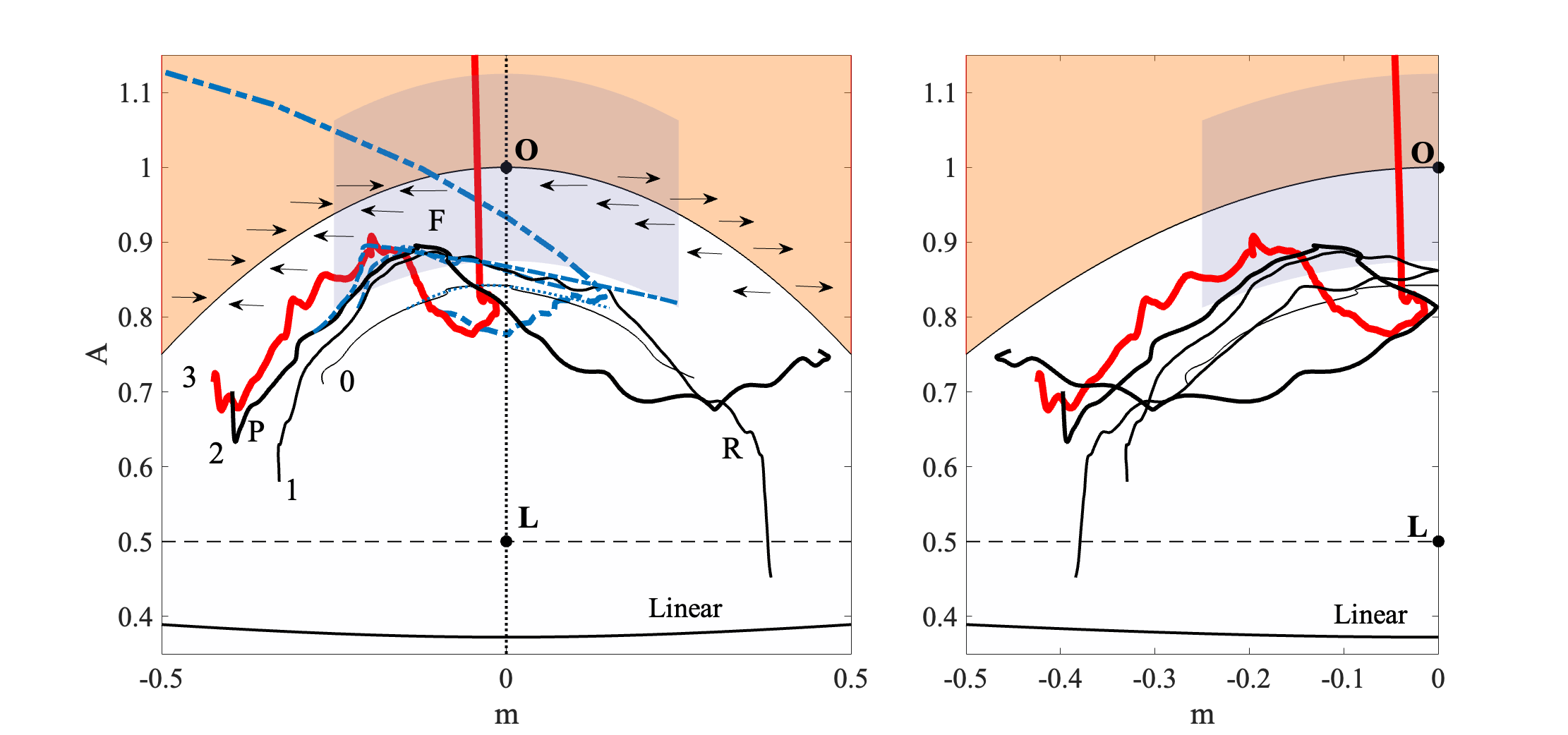}\\[1ex] 

  \caption{Fold catastrophe for inviscid and irrotational waves in deep water. (Left) fold boundary $\partial\mathcal{F}_0$ in the $(m,A)$ plane and trajectories (solid lines) of the four major successive waves of the simulated \emph{right-going} wave group $C3N5A0.514$~\citep{BarthelemyJFM2018}. These are labeled as $0,1,2,3$ at the beginning~(early time) of the trajectories, and depicted with increasing thickness. The thickest (dark red) line indicates the largest breaking wave. Points $P$, $F$, and $R$ mark the pre-focusing, focusing, and post-focusing phases of the largest non-breaking wave~($2$) (see Fig.\ref{FIG2waveprofiles+crestgroupspeeds}). Theoretical predictions (dashed lines) from the slow-fast model are also depicted. The path of a linear packet is also reported~(amplitude $\mu = k a = 0.1$ and spectral bandwidth $\nu = 0.3$). (Left) the desymmetrized half-domain~$m\le0$.}\label{FIG3fold}
\end{figure}

\subsection{Fold Catastrophe and Breaking Onset}
The equilibria of the surface slope equation in Eq.~\eqref{ODEdim4b} are the loci of the parabola $F(m,A)=- m^2 + 1 - A=0$, the boundary~$\partial\mathcal{F}_0$ of a fold in the the plane~$(m,A)$. Their stability depends on the sign of $F' =\partial_m F=-2 m$. An equilibrium at $m=m_0$ is stable if $F'(m_0;s,A)<0$, otherwise unstable if $F'(m_0;s,A)>0$. If $F'(m;s,A)=0$  the fixed point is marginal, neither stable or unstable. 

The fold boundary~$\partial\mathcal{F}_0$ is shown in the left panel of Fig.~\ref{FIG3fold} and the stable and unstable manifolds of equilibria are depicted with arrows. Equilibria with negative~(positive) slope are stable~(unstable). These equilibria coalesce at the tip $O$ at $(A=1, m=0)$, forming a marginal state with slope $m=0$. Thus, $\partial\mathcal{F}_0$ is normally hyperbolic, except at $O$. This boundary is also symmetric with respect to $m$, reflecting the parity symmetry of the Euler equations in deep water, which allows the same wave to propagate in either direction. In such cases, only the horizontal velocity~$U$, its gradient~$U_z$ and the hotspot speed $C_e$ change sign, leading to a corresponding sign change in both $s$ and the slope~$m$, except the focusing parameter~$A$. The symmetry can be removed by applying symmetry reduction techniques described in~\cite{ChaosBook}~(see section~\ref{paritysymmetry} and right panel of Fig.~\ref{FIG3fold} that displays the desymmetrized domain).  

The point $O$ is the marginal point of a fold catastrophe~\citep{Thom1975,zeeman1976catastrophe,arnold1984catastrophe}. Indeed, we can rewrite the surface slope equation as the gradient flow 
\begin{equation}\label{gradientfoldeq}
   \frac{dm}{d\tau} = -\frac{\partial V}{\partial m} \,,  
\end{equation}
where the cubic potential
\begin{equation}
    V(m,A)=(1-A)m -\frac{m^3}{3}\,,
\end{equation}
defines a cubic fold surface~\citep{Thom1975,arnold1984catastrophe}. 

The right panel of Fig.~\ref{FIG3fold} illustrates the trajectories (blue lines) in the plane~$(m,A)$ of the four major successive waves in the simulated \emph{right-going} wave group~$C3N5A0.514$ reported by~\cite{BarthelemyJFM2018}. These are labeled as $0,1,2,3$ at the beginning~(early time) of the trajectories. Line thickness increases with wave amplitude, with the thickest trajectory corresponding to the breaking wave~($3$).  Points~$P$, $Q$, and~$R$ denote the pre-focusing, focusing, and post-focusing phases of the largest non-breaking wave~($2$), whose profiles are shown in Fig.~\ref{FIG2waveprofiles+crestgroupspeeds}.
For comparison, the trajectory of a linear Gaussian packet is also included (wave steepness $\mu = ka = 0.1$, spectral bandwidth $\nu = 0.3$).

All trajectories of the four nonlinear waves, during their growing phase, are attracted toward the point~$O$ along the stable manifold of the negative-slope equilibria, and the energetic hotspot lies just behind the wave crest. They follow the parabolic arc toward the point~$O$, where the wave crest and the energetic hotspot coincide. The breaking wave~($3$) is the only one that crosses the boundary near~$O$, thereby entering its interior and undergoing finite-time blowup, indicating the onset of breaking. The numerical simulations, however, could not resolve the dynamics in the immediate vicinity of, or beyond, breaking. In contrast, the trajectories of the other non-breaking waves are repelled along the unstable manifold of the positive-slope equilibria, leading to the defocusing phase of the crests. A similar behavior is observed in other simulated breaking wave groups~\citep{BarthelemyJFM2018}.

By contrast, the linear packet remains far from the fold boundary, with $A \leq 1/2$, since $\epsilon \gg 1$ because $U_z$ is of the order of the small wave amplitude. The locus of inflection points that lies further away, at $A=1$, and is dynamically irrelevant for the energetic hotspot is not shown in the figure. Instead, the marginal point~$O$, where the wave crest coincides with the energetic hotspot, indicates that the dynamics observed from the two points are intrinsically linked to the onset of breaking~(see Fig.\ref{FIG2waveprofiles+crestgroupspeeds}). 

The slow–fast model captures the qualitative behavior of the wave dynamics near the breaking onset and can be further refined by extending the perturbation expansion to higher order. In practice, the model describes the wave dynamics qualitatively well even beyond its formal range of validity. Finally, higher-resolution simulations are needed to better resolve the wave field near the onset of breaking, particularly as the wave crosses the fold boundary near the  marginal point~$O$.


\subsection{Finite time blowup and breaking} 
Solving~\eqref{ODEdim4b} with initial conditions $m(\tau=\tau_0)=m_0$ yields the time evolution of the surface slope before the breaking onset as
\begin{equation}\label{mt}
    m(\tau) =\sqrt{1-A} \tanh\left [ \sqrt{1-A}\, (\tau-\tau_0) +\text{arctanh}\left(\frac{m_0}{\sqrt{1-A}}\right)\right]\,,\qquad \tau<\tau_c\,,
\end{equation}
where $A$ is a slowly varying adiabatic parameter. The surface slope~$m$ can blow up in a finite time after the wave crosses the turning point $A=1$ of the fold catastrophe at the later time $\tau_c>\tau_0$ with a flat surface. Then, when $A\ge1$, 
\begin{equation}\label{mt2}
   m(\tau)=-\sqrt{A-1} \tan\left [ \sqrt{A-1}\, (\tau-\tau_c)\right]\,,\quad \tau\ge \tau_c\,,
\end{equation}
and the breaking time is
 \begin{equation}
     T_{b}=\tau_b-\tau_c=\frac{\pi}{2 \sqrt{A_b-1}}\,,
 \end{equation}
where $A_b$ is the value at breaking. 
Fig.~\ref{FIG3fold} indicates a good qualitative agreement between the simulated surface slopes and the predicted counterparts based on Eq.~\eqref{mt}, where the latter are computed by inputting the numerically observed slow evolution of $A$. 

The extension of the region of validity of the fast–slow model~\eqref{ODEdim4b} beyond the breaking region, following the path~$\Gamma_*$ of Eq.~\eqref{gamma_m}, to include higher-order effects on the surface-slope evolution, requires a multiple-scale perturbation approach to solve for the slow dynamics and for the dynamics of the Hessians of the pressure and velocity fields in order to evaluate the hotspot speed~$C_e$ in Eq.~\eqref{Ce2} away from the marginal point~$O$. Then, the fold catastrophe could be the cross-section of a cubic cusp catastrophe, or a higher-order swallowtail type~\citep{Thom1975,arnold1984catastrophe,zeeman1976catastrophe}.

\begin{figure}[htbp]
  \centering

    \includegraphics[width=0.9\linewidth]{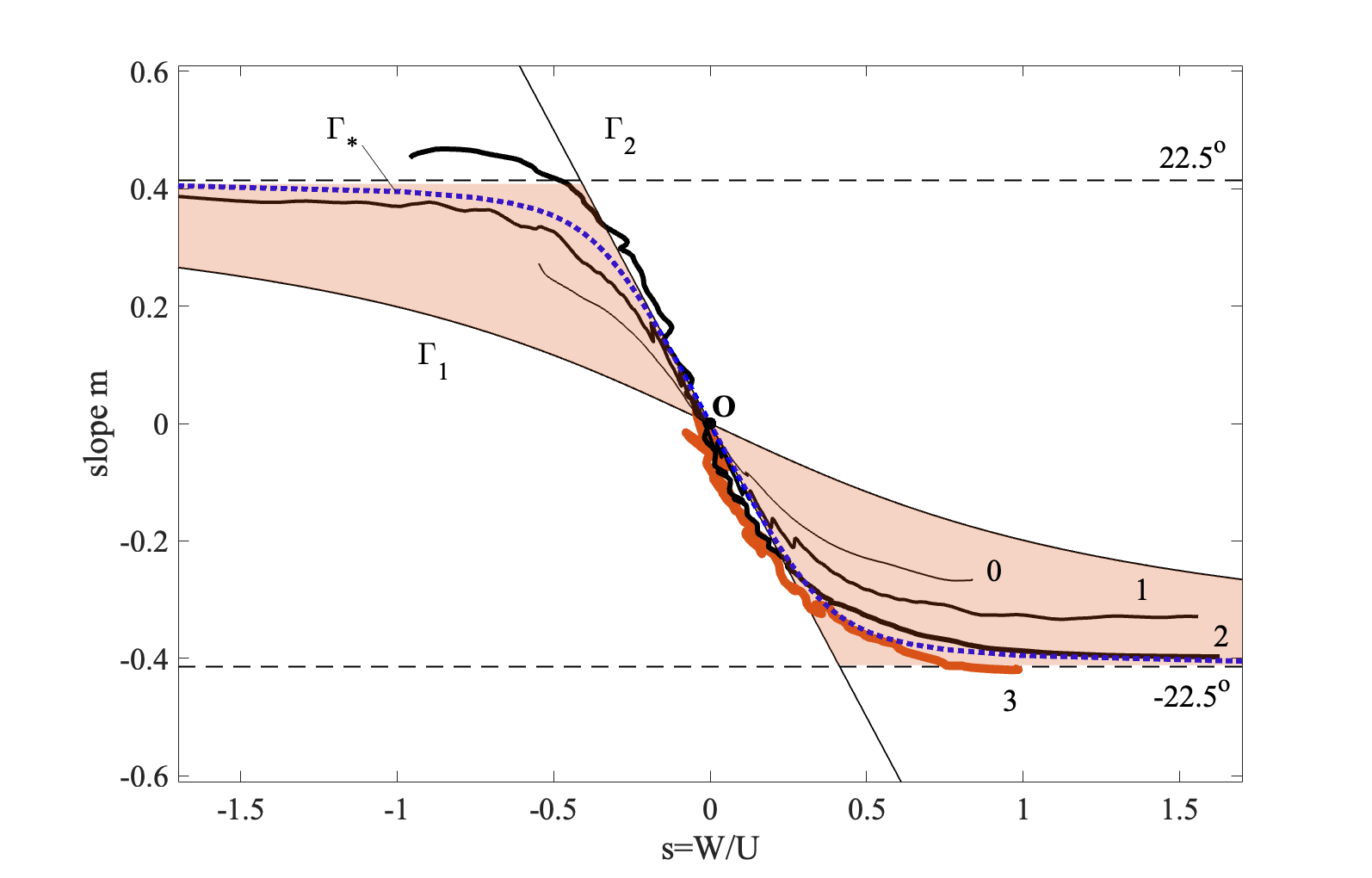}

  \caption{Wave dynamics in the $(s,m)$ domain. Trajectories (black lines) of the four major successive waves in the simulated \emph{right-going} wave group $C3N5A0514$, reported in~\cite{BarthelemyJFM2018}, are shown. These trajectories are labeled $0,1,2,3$ at their initial points (before focus) and are depicted with increasing thickness. The thickest (red) line corresponds to the breaking wave. The curves $\Gamma_1=\{(s,m):\tan\theta^+_{U} + m=0 \rightarrow s=4(-m+m^3)/(1-6m^2+m^4)\}$ and $\Gamma_2 =\{(s,m):m_2=0\rightarrow s=-m\}$, delimit the wave dynamics between the asymptotic slopes $\theta_*=\pm22.5^{\circ}=\pm\arctan(\sqrt{2}-1)$. The curve $\Gamma_*={(s,m):s=(-m+5m^3)/(1-6m^2+m^4)}$ represents the approximate maximally non-breaking (marginal) wave. The breaking wave (red) exceeds the $\theta_*$ threshold, marking the onset of breaking inception.
  }\label{ms_domain}
\end{figure}

\begin{figure}[htbp]
  \centering
  
\includegraphics[width=1.05\linewidth]{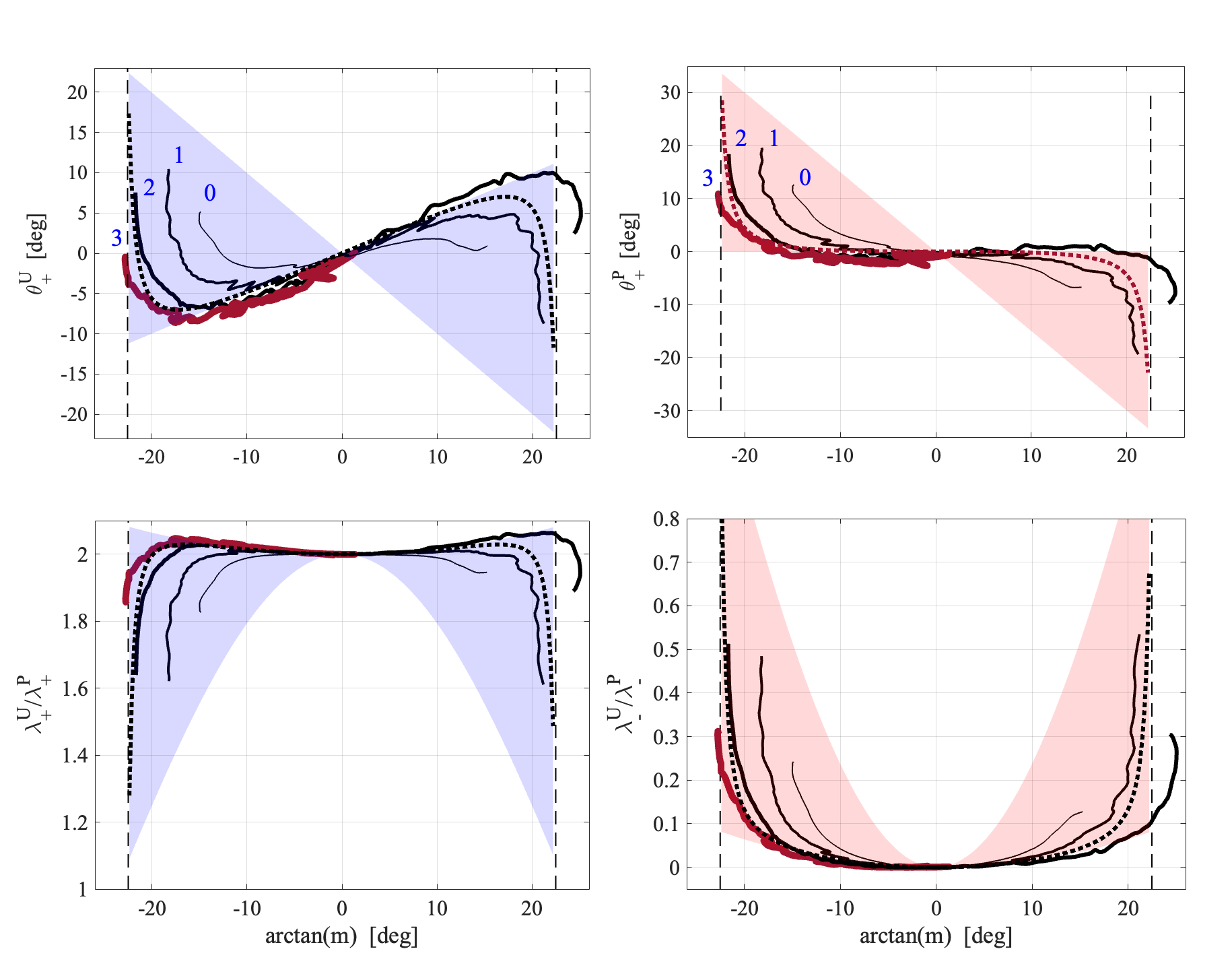}\\[1ex]  
 
  \caption{Eigen-structure of the weight matrices  $G_u$ and $G_p$ associated with the Hessians of the fluid velocity~$U$ and pressure~$P$, respectively.
Top panels: (Left) slope angle~$\theta^U_+$ of the eigenvector of the velocity weight matrix $G_u$ corresponding to the positive eigenvalue~$\lambda^U_+$ as a function of the surface slope~$m$, for the four major successive waves of the simulated \emph{right-going} wave group $C3N5A0514$~\citep{BarthelemyJFM2018}. (Right) Same, but for the slope angle~$\theta^P_+$ of the eigenvector of the pressure weight matrix $G_P$ associated with the positive eigenvalue~$\lambda^P_+$. The breaking wave is shown as the thickest dark-red line.
Bottom panels: (Left) Ratio $\lambda_+^U/\lambda_+^P$ of the positive eigenvalues, and (right) ratio $\lambda_-^U/(-\lambda_-^P)$ of the negative eigenvalues.
The waves are labeled $0,1,2,3$ at the beginning (before focus) of their trajectories and are depicted with increasing thickness. The breaking wave is the thickest curve, colored distinctly from the non-breaking waves. The non-breaking waves are confined within the shaded region bounded by two limiting cases: (i) the condition $\tan\theta^U_{+}+m=0$, and (ii) vanishing momentum parameter $m_2=0$~(curves~$\Gamma_1$, $\Gamma_2$ in Fig.~\ref{ms_domain}).  The dotted line represents the approximately maximally non-breaking (marginal) wave~(curve~$\Gamma^*$ in Fig.~\ref{ms_domain}). The breaking wave surpasses the region bounded by the surface angle~$\theta_*=\pm22.5^{\circ}=\pm\arctan(\sqrt{2}-1)$ (dashed lines), marking the inception of breaking.
  }\label{lambda_eig}
\end{figure}

\subsection{The point of no return: breaking inception}\label{Sec:Point of no returm}
The breaking inception marks the point of no return, beyond which waves irreversibly break through a fold bifurcation, trespassing the the boundary~$\partial \mathcal{F}_0$ at the onset of breaking (see Fig.~\ref{FIG3fold}). This  boundary defines a skeleton in the domain $(m,A)$ that guides the wave dynamics near breaking. Interestingly, the stability of equilibria, expressed via $\partial_m F$, does not depend on $A$, which serves to define the  fold boundary. 
Moreover, the point of breaking inception is robust and independent of how the waves are generated~\citep{DerakhtiJGR2020,BoettgerJFM2023,BoettgerPRF2024}.
In~\ref{C_e_derivation} we have derived the energetic hotspot speed that follows from Eq.~\eqref{Ce2} as 
\begin{equation}\label{Ce2a}
C_e = U + \frac{ \mathrm{trace}(G_p\,\mathcal{H}(p)\big|_Q)/\rho+\frac{(1 + m^2) ( g + (1 + m s ) P_z/\rho )}{U(1 - m s)}U_z -\frac{ (1+m^2) (1+m s) (s^2+1) }{(1- m s)^2}U_z^2 }{\mathrm{trace}(G_u\,\mathcal{H}(u)\big|_Q) +\frac{(s^2+1)\eta_{xx}|_Q  }{1- m s}U_z+\frac{(m^2+1)^2 (s^2+1) }{U(1- m s)^2}U_z^2 }\,,
\end{equation}
where the trace of a field $f$
\[
\mathrm{trace}(G_f\,\mathcal{H}(f))=\sum_{ab} g_f^{ab} H_{ab}(f)\,,
\]
and the Hessian matrices~$\mathcal{H}(p),\mathcal{H}(u)$ of pressure and horizontal fluid velocity fields are evaluated at the energetic hotspot~$Q$, and encode the second-order curvature effects of the fields. Note that the trace of $\mathcal{H}(u)$ is null because $u$ is a harmonic function. 
The symmetric $2\times 2$ tensors are given by
\[
G_u=[g_u^{ab}]  = 
\begin{pmatrix}1-ms & \frac{1}{2}(s+2m-m^2s)\\[1mm] \frac{1}{2}(s+2m-m^2s) & m(m+s)\end{pmatrix}\,,
\]
with determinant~$g_u= -s^2(1 +m^2)^2/4$,
and
\[
G_p=[g_p^{ab}] =
\frac{1}{2} 
\begin{pmatrix} 1-ms & m+s\\[1mm] m+s & m s -1\end{pmatrix}\,,
\]
with determinant~$g_p=-(1+m^2)(1+s^2)/4$. These matrices encode the anisotropy of velocity and pressure fields  and their effects on the wave surface dynamics. In particular, they weight the contribution of the two Hessians through their eigenvalues and eigendirections. Since the matrices are symmetric, their respective eigenvectors are also orthogonal. 
Their inverses are the matrix representations of pseudo-Riemannian, specifically \emph{hyperbolic} metrics encoding anisotropy~(see~\ref{C_e_derivation}).  

\begin{figure}[htbp]
  \centering
  \hspace*{-0.5cm}\includegraphics[width=1.05\linewidth]{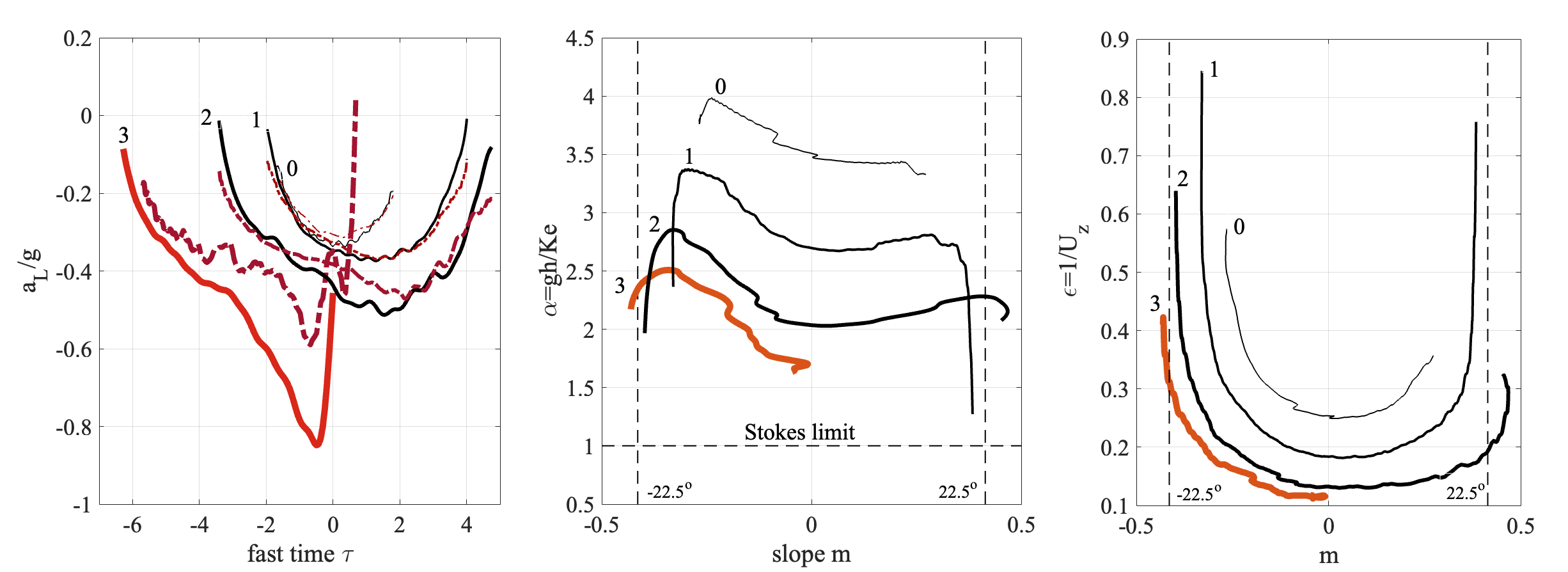}
  
  \caption{(Left) Vertical Lagrangian acceleration $a_L/g$ (solid lines) and theoretical predictions from the slow-fast model (dashed lines) at the energetic hotspot as function of the fast time~$\tau$, (center) ratio $\alpha=g h/K_e$  and (right) $\epsilon=1/U_z$ versus the surface slope $m$ for the four successive waves of the simulated \emph{right-going} wave group $C3N5A0514$~\citep{BarthelemyJFM2018}. These trajectories are labeled $0,1,2,3$ at their initial stages (before focus) and are depicted with increasing thickness. The thickest (red) line corresponds to the breaking wave. At the breaking inception, the surface slope corresponds to the angle~$\theta_*=-22.5^{\circ}$. For the limiting steady Stokes wave~$\alpha=1$. 
  }\label{ghKe_fig}
\end{figure}

Near the breaking onset, where $\epsilon=1/U_z\ll1$, from the scalings in Eqs.~\eqref{scalinga},\eqref{scaling2a}, the energetic hotspot speed~$C_e$ is nearly the same as the equivalent crest speed~$C_{req}$, and the Hessian terms are negligible, that is
\[
C_e=U + \frac{U_z}{\eta_{xx}|_Q}+O(\epsilon^{3/2})\,.
\]
However, away from the breaking onset, especially near the inception the two Hessian terms cannot be neglected, since the wave leading to break is still growing and $\epsilon$ is not small, as clearly seen in the right panel of Fig.~\ref{ghKe_fig}.   


To see this, in the plane~$(m,s)$ we identify two extreme paths that delimit the dynamics of non-breaking waves, as illustrated in Fig.~\ref{ms_domain}. Along the path
\begin{equation}\label{gamma_1}
\Gamma_1=\left\{(s,m): \tan\theta^+_{U} + m=0\rightarrow s=4\,\frac{-m + m^3}{1 - 6 m^2 + m^4}\right\}\,,
\end{equation}
the slope of the eigenvector of the fluid velocity matrix $G_u$ corresponding to a positive eigenvalue  equals minus the surface slope~$m$. Along the path
\begin{equation}\label{gamma_2}
    \Gamma_2=\left\{(s,m): m_2=0\rightarrow s=-m\right\}\,,
\end{equation}
the pressure matrix~$G_p$ is diagonal and the momentum parameter $m_2$ vanishes, implying vertical transport of horizontal momentum density towards the surface dominating over the horizontal advection. The curve $\Gamma_1$ has two asymptotes where the denominator $1 - 6 m^2 + m^4=0$, corresponding to surface slopes~$\pm m_*=\pm (\sqrt{2}-1)$ and its orthogonal complement $\pm(1+\sqrt{2})$. The corresponding angle is~$\theta_{*}=\pm \arctan{(\sqrt{2}-1)}=\pm 22.5^{\circ}$ and its orthogonal complement $\pm 67.5^{\circ}$. The region of validity for $\Gamma_1$ is within $|m|\le m_*$. The path~$\Gamma_*$ approximates the maximally steep non-breaking wave (nearly wave~$2$), defined as
\begin{equation}\label{gamma_m}
    \Gamma_*=\left\{(s,m): s=\frac{-m + 5m^3}{1 - 6 m^2 + m^4}\right\}\,.
\end{equation}
Figure~\ref{ms_domain} shows the dynamics of the four major successive waves of the simulated \emph{right-going} wave group $C3N5A0514$~\citep{BarthelemyJFM2018}. The non-breaking waves are confined within the area bounded by the two extreme paths~$\Gamma_1$ and $\Gamma_2$, with the path~$\Gamma_*$ representing the approximate maximally steep non-breaking (marginal) wave, close to wave~$2$. The breaking wave~$3$ trespasses the region delimited by the slope~$\theta_{*}=\pm 22.5^{\circ}$ (dashed lines), which defines a breaking inception. 
Figure~\ref{lambda_eig} summarizes the eigenstructure of the matrices $G_u$ and $G_p$ associated with the pressure~$p$ and velocity~$u$ Hessians tracked at the energetic hotspot. The point of breaking inception marks the maximal anisotropy between the velocity and pressure fields that the wave can sustain. This marks the dominance of the velocity Hessian at the breaking inception. 

The role of the velocity and pressure Hessians in determining the breaking inception has a close analogue in turbulence. In incompressible flows, the pressure field is governed by a nonlocal Poisson equation, and its second derivatives enter the evolution equations for the velocity gradient tensor via the pressure Hessian~\citep{ohkitani1995pressureHessian,Chellard2006pressureHessian}. 
The anisotropic part of the pressure Hessian prevents finite-time singularities by counteracting excessive alignment of strain and vorticity. Similarly, in the wave breaking problem, the weighted Hessians of pressure and velocity fields rule the approach to breaking.

Breaking depends on the geometric nature of second-order velocity and pressure information: while the first derivatives describe instantaneous shear, stretching and pressure gradients, the Hessian encodes how these vary in space. The associated weight matrices encode anisotropic pathways through which the fold catastrophe is either suppressed or allowed. 

This trend is robust, as it consistently appears across other simulated breaking wave groups in~\cite{BarthelemyJFM2018}, indicating that the observed wave behavior and breaking inception is not case-specific but reflects a fundamental feature of the wave dynamics.

\subsection{The $B_+$ and $B$ parameters}

To characterize the inception of breaking at the energetic hotspot, \cite{BoettgerJFM2023} introduced the parameter
\begin{equation}\label{Bpar}
B_+ = \frac{\lVert \mathbf{U} \rVert}{\lVert \mathbf{V}_Q \rVert} = \sqrt{\frac{U^{2} + W^{2}}{C_e^{2} + v_Q^{2}}} = B \sqrt{\frac{s^{2} + 1}{1 + [m + B (s - m)]^{2}}}\,,
\end{equation}
where $\mathbf{U}$ is the Eulerian velocity vector, $\mathbf{V}_Q$ is the velocity of the energetic hotspot~$Q$, and $B = U / C_e$ is the ratio of their horizontal components. From the kinematic condition~\eqref{kin}, $v_Q = C_e m + U (s - m)$. 

At the energetic inception, we find $B_+^*\approx0.6$~($B\approx0.55)$ with $s^*\approx1$ and the surface slope $m^*=\sqrt{2}-1\approx 0.41$ both still significant away from the breaking onset~(see Fig.~\ref{FIG2_fastslow}). The momentum parameter $m_2^*=-1$ suggesting the vertical momentum transport is not dominant, while horizontal advection flows momentum toward the crest region. \cite{BoettgerJFM2023} also found that eventually the value of $B_+$ passes through $0.85$ at some time later prior to the breaking onset as the energetic hotspot gets closer to the crest. That is the same breaking inception threshold as observed by following the crest~\citep{BarthelemyJFM2018}. Thus, the marginal point~$O$ in Fig.~\ref{FIG3fold}, where the wave crest coincides with the energetic hotspot, suggests that the dynamics observed from these two points are intrinsically linked to the onset of breaking. In particular, the coincidence of crest and hotspot motion at~$O$ marks the onset to breaking: the wave transitions from reversible focusing dynamics to irreversible breaking via a fold catastrophe~\citep{Thom1975,zeeman1976catastrophe,arnold1984catastrophe}.

Estimating the critical threshold~$B_+^*$ from Eq.~\eqref{Ce2} is a non-trivial task. It requires extending the region of validity of the fast-slow model~\eqref{ODEdim4b} beyond the breaking region~(see Fig.~\ref{FIG3fold}). Nevertheless, we can derive lower bounds as follows. 
At the breaking inception, the wave is still in the growing phase, and the hotspot elevation $h$ satisfies $\mathrm{d}h/\mathrm{d}\tau \ge 0$ (see Eq.~\eqref{crestheight} in~\ref{appAh}). This yields the lower bound $B^*_+ \ge m^*/(m^*-s^*)\approx 0.3$.

\subsection{Vertical Lagrangian acceleration}
The vertical Lagrangian acceleration at the tracking point~$Q$ follows from Eq.~\eqref{verta_L} as
\begin{equation}\label{verta_L2}
    a_L = -(P_z/\rho + g ) = U U_z \left(A-1 - \beta +m^2\right)\,, 
\end{equation}
where the term between square parentheses is $O(\epsilon)$ and balances $U_z=1/\epsilon$ to produce a finite Lagrangian acceleration at breaking.
The left panel of Fig.~\ref{ghKe_fig} depicts the vertical Lagrangian acceleration $a_{L}$ inferred from simulations. Note that it exceeds the Stokes limit $g/2$, and in fair agreement with the predicted theoretical acceleration $a_L$ in Eq.~\eqref{verta_L2}, with the exception of underpredicting the breaking case. The simulations suggest that $a_L\rightarrow-g$ as the wave approaches the breaking onset. This critical threshold was originally proposed by \cite{phillips1966dynamics}, and was later shown by \cite{Bridges_2009} to mark the transition from elliptic to hyperbolic character of the Euler equations. Higher-resolution simulations are needed to better resolve the wave field and surface to prove or disprove that $a_L+g\rightarrow0$ as breaking is approached.

\subsection{Maximum Crest Height}
Consider the dimensionless parameter tracked at the energetic hotspot~$Q$ 
\begin{equation}
   \alpha = \frac{g\, h}{K_e}=\frac{C_{k_e}}{U}-1\,,
\end{equation}
where $g$ is the gravitational acceleration, $h$ is the surface elevation above the mean water level, and $K_e = (U^2 + W^2)/2$ denotes the local kinetic energy density of the fluid motion at~$Q$. From Eq.~\eqref{Kefluxspeed}, $\alpha$ measures the horizontal kinetic energy flux  velocity~$C_{k_e}=U(g h + K_e)$ relative to the fluid speed at the wave surface. 

From Bernoulli’s equation~\eqref{Bernoulli} on the free surface,
\begin{equation}\label{Bernoulli_surface}
g h + K_e + \partial_t \phi\big|_Q = 0 \,, \qquad z=\eta(x_Q,t)\,,
\end{equation}
we obtain
\begin{equation}
\alpha = -\left(1 + \frac{ \partial_t \phi\big|_Q }{K_e}\right)\,,
\end{equation}
where $\phi$ is the velocity potential. Since $\alpha>0$ at a wave crest, the dynamic pressure term $\partial_t \phi|_Q$ must be negative and larger in magnitude than the kinetic component $K_e$, that is, $\partial_t \phi|_Q < -K_e$.  

For the limiting steady Stokes wave, $\partial_t \phi|_Q = -2 K_e$ and $\alpha = 1$, while the vertical Lagrangian acceleration $a_{L}^{Stokes} = -U U_z= -P_z - g = -g/2$, with $P_z = -g/2$ being the vertical pressure gradient at the crest~\citep{Longuet_Higgins_AccelerationsinSteepGravityWaves,Shemer2013nhess}.  
For nonstationary waves, however, $\partial_t \phi$ is expected to exceed the steady Stokes value in magnitude, reflecting the enhanced local accelerations of the fluid near steep, evolving crests.

The  center panel of Fig.~\ref{ghKe_fig} shows the hysteresis loops of $\alpha$ versus the surface slope~$m$ for four successive waves of the simulated \emph{right-going} wave group $C3N5A0514$~\citep{BarthelemyJFM2018}. The ratio $\alpha$ is significantly greater than the Stokes limit for all cases and is bounded from below by $\alpha^* \approx 1.6$, indicating that nonstationarity plays a dominant role near breaking. This corresponds to the maximum surplus of kinetic energy~$K_e$ over potential energy~$g h$ that a wave can sustain, beyond which breaking becomes inevitable. This trend is observed in other simulated wave groups, suggesting upper bounds on the crest height of oceanic rogue waves~(see, e.g., \cite{Fedele2025RogueWaves} and references therein). Higher resolution numerical simulations are required to confirm these bounds and the full evolution of waves approaching the onset of breaking and beyond. 

\subsection{Finite-amplitude steady Stokes waves}\label{standingwaves}
 Consider the left panel of Fig.~\ref{FIG3fold}. The point $L$ at $(m=0, A=1/2)$ in the plane~$(m,A)$ can be associated with a linear steady Stokes wave having group velocity $C_g = C_0/2$, where $C_0$ is the phase speed. If the energetic hotspot is taken to coincide with the crest of the Stokes wave, and its speed $C_e$ is identified with $C_g$, then it follows that $A=1/2$ and $m=0$ at the crest of a linear Stokes wave. 

A linear wave group can thus be regarded as induced by a perturbation of the linear Stokes wave at~$L$ by spectral broadening. We note that \cite{DeconinckStokes2023} induced breaking by perturbing finite-amplitude nonlinear Stokes waves. This suggests that the segment between $L$ and $O$ could represent the locus of nonlinear Stokes waves, with the limiting Stokes wave located at~$O$.  

If the energetic hotspot speed $C_e$ is defined as the energy flux velocity of the finite-amplitude Stokes wave~\citep{PeregrineThomas1979, LonguetHiggins1975}, then \cite{Fenton1985}~fifth-order solution gives the finite-amplitude Stokes waves lie on $L-O$ emanating from $L$. However, proving that the highest (limiting) Stokes wave is at the tip~$O$ is not straightforward. Its group velocity tends to a finite value~\citep{LonguetHigginsFox1978part2}, but evaluating the limiting value of the velocity gradient~$U_z$ and $A$ requires refined numerical computations such as those performed by~\cite{Creedon_Deconinck_Trichtchenko_2022} and \cite{DeconinckStokes2023}. At present, therefore, the representation of finite-amplitude Stokes waves by the segment $L-O$, with the limiting Stokes wave at the marginal point~$O$ remains a conjecture to prove or disprove.

\subsection{Parity-reflection symmetry}\label{paritysymmetry}
Geometrically, the vertical axis $m=0$ serves as a parity-symmetry line of the fold boundary~$\mathcal{F}_0$. When crossing this axis, the stability of equilibria flips: stable (unstable) fixed points become unstable (stable).  Such a parity-reflection symmetry may be removed by applying symmetry reduction techniques described in~\citep{ChaosBook}. For instance, for waves propagating to the right~(left), the trajectory $m(t), A(t))$ can be restricted to the half-domain $m < 0$~($m>0$). Whenever it crosses the line $m = 0$, the signs of $m$ is flipped, thereby folding the trajectory back into the $m<0$~($m>0$) half-domain~(see right panel of Fig.~\ref{FIG3fold}).

\begin{figure}[htbp]
  \centering
  \includegraphics[width=0.75\linewidth]{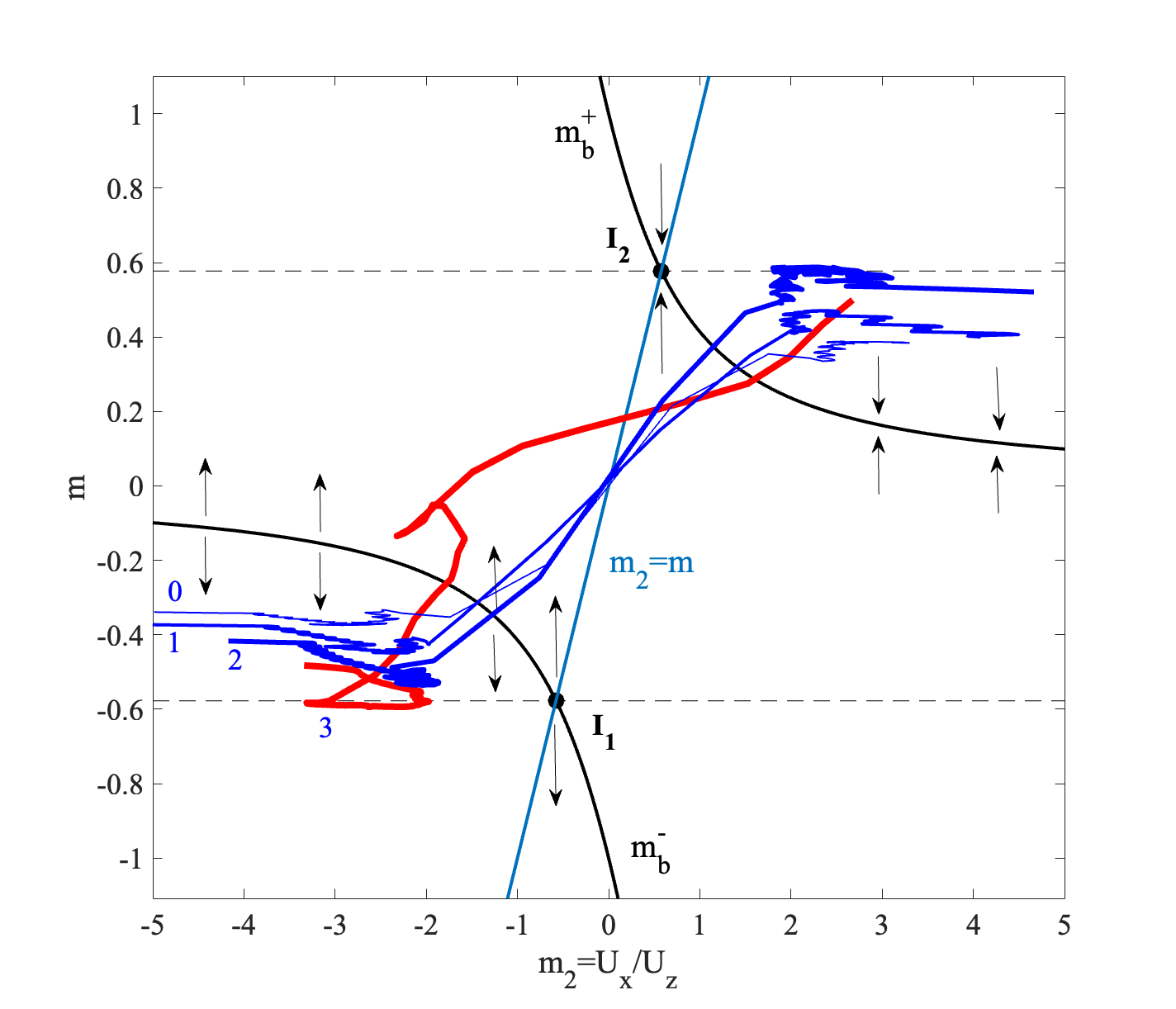}\\[1ex] 
  
\caption{Dynamics in the plane~$(m_2,m)$ as observed from the inflection point. The points $I_1$ and $I_2$ denote the Stokes limiting slopes $m_b = \mp\tan(\pi/6)$ for right-going~($-$) and left-going~($+$) waves, respectively.  The thick black lines indicate the loci of stable and unstable fixed points of  the slope equation~\eqref{slopeIP}. Trajectories (blue lines) of the four tallest successive waves of the 
\emph{right-going} simulated wave group $C3N5A0.514$ reported in~\cite{BarthelemyJFM2018}  are shown with increasing thickness, with the thickest (red) line corresponding to the breaking wave. 
See also Figure~\ref{FIG2waveprofiles+crestgroupspeeds}, which depicts the surface  and energy profiles of wave~$2$ and the location of its inflection point.}
\label{FIG2D_m2_m_inflectionpoint}
\end{figure}

\section{Wave breaking following the inflection point}
\cite{McAllister2023} performed numerical simulations of $2$D inviscid and irrotational water waves without surface tension. By tracking the surface point of maximum inflection, where $\eta_{xx} = 0$, they found that the minimum local interface slope $m_b=\tan(\pi/6) \approx 0.5774$ at that point separates breaking and non-breaking waves. This threshold corresponds to an angle of $30^\circ$, matching the limiting steepness derived by Stokes~\cite{Stokes1847}. Such a threshold is not robust, unlike the one following the energetic hotspot. \cite{Boettger2025} pointed out that surface tension or wind stresses act as additional forces along the curved air–water interface, affecting the critical slope $m_b$ that increases to nearly~$60^{\circ}$.  Note that the inflection point of a focusing right~(left)-going wave is located on the front~(back) face of the wave, and its slope is negative~(positive) decreasing~(increasing) as the wave steepens.

From Eq.~\eqref{ODE2} in~\ref{appAODE}, the surface slope~$m$ tracked at the inflection point evolves according to the ODE: 
\begin{equation}\label{slopeIP}
\frac{dm}{d\tau} = -2\, m_2 m + (1 - m^2) \,.
\end{equation}
This explicitly depends on the momentum parameter~$m_2$ defined in Eq.~\eqref{m2eq}. From~\eqref{slopeIP}, there are two loci, or branches, of slope equilibria, 
\[
m_b^{\pm} = -m_2 \pm \sqrt{1 + m_2^2}\,.
\]
The locus $m_b^{-}$ is always unstable, whereas $m_b^{+}$ is always stable, independent of the value of $m_2$, as illustrated in Fig.~\ref{FIG2D_m2_m_inflectionpoint}. The two roots are distinct since the discriminant $\Delta = 4(m_2^2+1)$ never vanishes. The line $m_2=m$ intersects the two branches at the points $I_1$ and $I_2$, which correspond to the Stokes limiting slopes $\mp m_b$ for right-going~($-$) and left-going~($+$) waves, respectively.  In the same figure, we show the trajectories (blue lines) of the four tallest successive waves extracted from the numerical simulation of the \emph{right-going} 
unidirectional focusing group $C3N5A0.514$ reported in~\cite{BarthelemyJFM2018}. The surface and energy profiles of wave~$2$, together with the location of its inflection point, are displayed in Fig.~\ref{FIG2waveprofiles+crestgroupspeeds}. 

The four trajectories follow the stable and unstable directions of the dynamical flow; however, a clear breaking inception at $I_1$, directly associated with a cusp catastrophe, is not evident. However, breaking occurs for wave~($3$) as the surface slope exceeds the Stokes limit~$m_b$, in agreement with~\cite{McAllister2023}. At the breaking inception~($I_1$), the momentum parameter takes the value $m_2 = -m_b \approx -0.5774$, which is negative and comparable with $m_2^*=-1$ observed from the energetic hotspot. Thus, the vertical momentum transport is not dominant at the inflection point, and horizontal advection transports momentum toward the crest region.  

The physical mechanisms leading to breaking inception at the critical Stokes slope can be identified by examining the horizontal speed $C_{ip}$ of the inflection point, derived in~\ref{appAnisotriopy}: 
\[
C_{ip}  = U + \frac{\sum_{ab} g^{ab} U_{ab}}{\eta_{xxx}|_Q}
= U + \frac{-2 m\, U_{xx} + m(1-m^2) U_{zz} + (1-3 m^2) U_{zx}}{\eta_{xxx}|_Q}\,.
\]
This speed depends sensitively on the anisotropy of the flow field, encoded in the Hessian of the velocity potential tracked at the inflection point~$Q$,
\[
\mathcal{H}(u)\big|_Q=
\begin{pmatrix}
U_{xx} & U_{xz}\\
U_{xz} & U_{zz}
\end{pmatrix}\,,
\]
and weighted by the symmetric contravariant tensor $g^{ab}$. Its inverse is the matrix representation of the following metric $g_{ab}$ in coordinates $(x,z)$:
\[
ds^2 = g_{ab}\, da\, db
= -\frac{4\, m(1-m^2)}{1 + 2 m^2 + m^4}\, dx^2
+ \frac{8m}{1 + 2 m^2 + m^4}\, dz^2
+ \frac{4(1-m^3)}{1 + 2 m^2 + m^4}\, dx\, dz\,.
\]
This metric is pseudo-Riemannian and \emph{hyperbolic}, since $g<0$, with eigenvalues of opposite sign:
\[
\lambda_\pm(m) = \tfrac{1}{2}(1+m^2)\left(-m \pm \sqrt{1+m^2}\right)\,. 
\]
Through these weights, the metric rules how anisotropic fluid deformations affect the evolution of~$C_{ip}$. While the velocity gradients $(U_x,U_z)$ describe local shear and stretching, their second derivatives $(U_{xx},U_{zz},U_{xz})$ measure the spatial variation of these rates, or inhomogeneity of the flow field. The eigenvectors of $g_{ab}$ identify the principal directions along which these anisotropic deformations are most effective. The two principal directions have angles with respect to the horizontal given by~$\theta_{\pm}=\pm\pi/4 + 3/2\arctan{m}$.

\begin{figure}[htbp]
  \centering
  \includegraphics[width=0.75\linewidth]{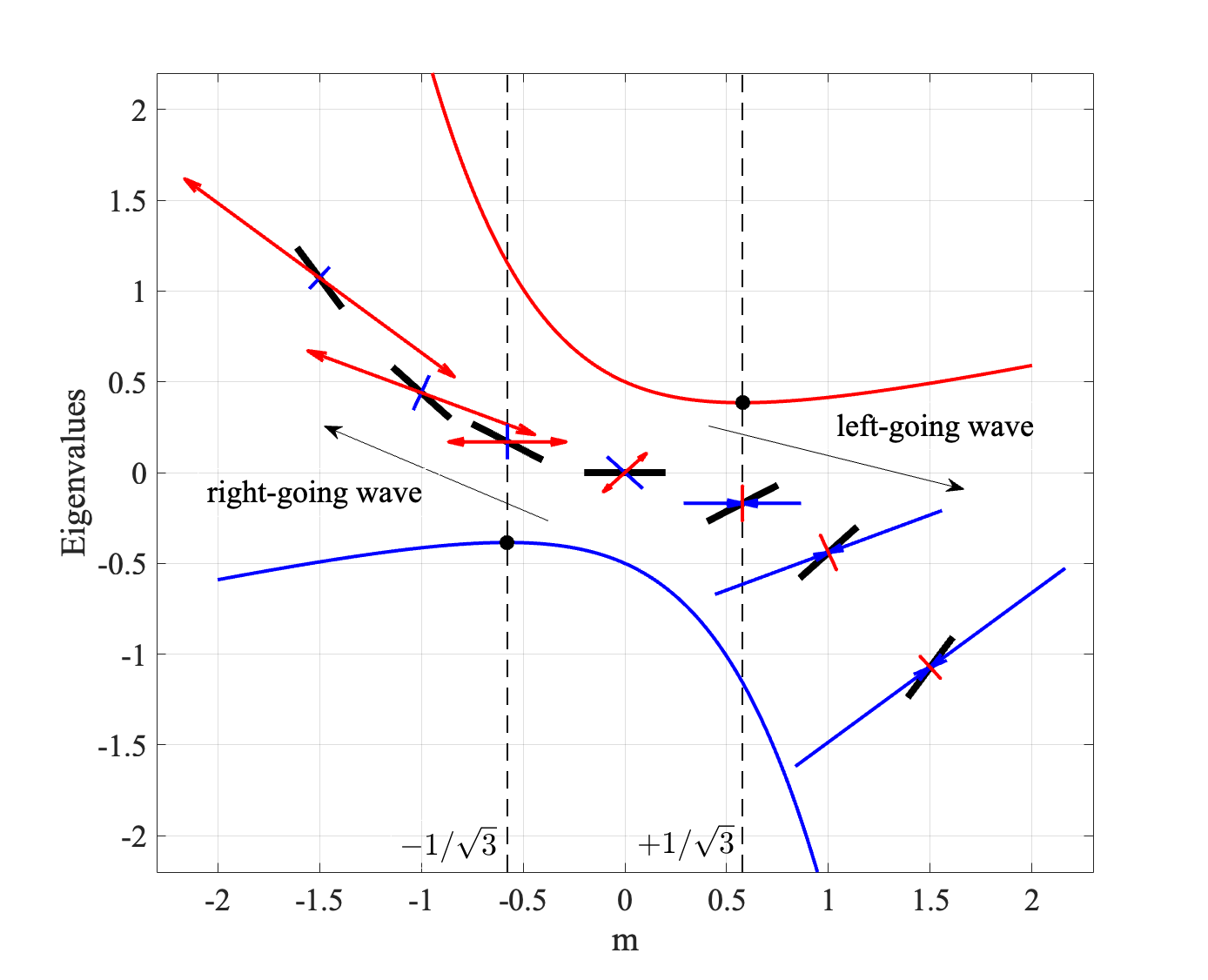}\\[1ex] 
  
\caption{Eigenvalues~$\lambda_+$~(red) and $\lambda_-$~(blue) of the metric~$g_{ab}$ as a function of the surface slope $m$ at the inflection point. The principal directions of the metric~$g_{ab}$ are depicted for values  $m=-1,-m_b,0,m_b,1$ of the surface slope~(thick black line). $m_b=1/\sqrt{3}$ is the limiting Stokes slope corresponding to a $30^{\circ}$ angle. The inflection point of a focusing right~(left)-going wave is located on the front~(back) face of the wave. Its slope is negative~(positive) and decreases~(increases) as the wave steepens.}
\label{principaldirectionsIP}
\end{figure}

In particular, the mixed coefficient $g_{xz}$ vanishes at the limiting Stokes slope $m=\pm m_b$, where the eigenvalues also exhibit local minima and maxima, as shown in Fig.~\ref{principaldirectionsIP}. Beyond this threshold, the anisotropy is enhanced: stretching becomes strongly biased along one principal axis that tilts to align along the steep surface, creating conditions favorable for wave overturning or spilling. This geometric mechanism highlights how wave breaking onset is tied to the interplay between velocity curvature and surface anisotropy at the inflection point.

\section{Black hole analogy and geometrization of wave breaking}\label{sec:BH}

We have observed that, as a wave approaches the onset of breaking, its effective dynamics slow down. This is analogous to matter approaching a non-stationary black hole~\citep{Vaidya1951}. A Vaidya spacetime is a generalized Schwarzschild-like black hole with time-dependent mass, capable of describing collapsing, accreting systems or radiating stars losing mass through light emission~(see, e.g.,~\cite{MisnerThorneWheeler1973,Carroll2004}). 

The standard gravity analogue is typically based on the characteristic speed of wave perturbation fields of a fluid flow, allowing for a spatially varying speed of sound through the medium~\citep{UnruhGravityanalogue2002,GravityAnalogue2011_Liberati}. In contrast, the analogy presented here does not rely on any perturbative fluid waves; instead, it emerges directly from the intrinsic, fully nonlinear dynamics of the energetic hotspot of the evolving wave surface. 

Consider the relative motion of the fluid in the Lagrangian frame moving with the horizontal velocity~$U$:
\[
dr = dx - U(x,t) dt\quad \rightarrow \quad r  = x - \int U(x,t) \,\mathrm{d}t\,.
\]
Then, the motion of the energetic hotspot can be thought as that in the radial sector of a 4D spherically symmetric metric in the coordinates $(t,r,\Omega)$: 
\begin{equation}\label{BHmetric}
    ds^2 = -\left(C_e(r,t)-U(r,t)\right)dt^2 + \frac{dr^2}{C_e(r,t)-U(r,t)} + r^2 d\Omega\,,
\end{equation}
where $C_e(\tilde{f}_1,\tilde{f}_2,\cdots)$ is the speed of the energetic hotspot defined in Eq.~\eqref{Ce2a}.
This speed depends intrinsically on $r$ and $t$ through the evolution of several scalar-field observables $\tilde{f}_j(r,t)$ measured on the wave surface, such as slope, velocity, and pressure gradients. In general, we write their evolution laws:
\begin{equation}
\partial_t \tilde{f}_j + (C_e - U)\,\partial_r \tilde{f}_j = F_j(\tilde{f}_1,\tilde{f}_2,\dots), 
\qquad j = 1,\dots,N,
\end{equation}
with explicit forms for specific observables given in~\ref{appA}. These fields play the role of effective matter sources that generate the geometry. From the Einstein equations one can formally associate an energy–momentum tensor to this geometry that should satisfy the covariant conservation laws. This condition ensures that energy–momentum is transported consistently along the geometry~\citep{MisnerThorneWheeler1973, Carroll2004}.

\subsection*{Null rays, apparent and event horizons}

For radial null rays~($d\Omega=0$), the null condition~$ds^2=0$ gives the slopes of ingoing $(-)$ and outgoing $(+)$ null geodesics
\begin{equation}
\frac{dr}{dt} = \pm \left(C_e - U\right),
\end{equation}
which are the trajectories of the energetic hotspot on the surface of right-going and left-going waves. This motion can be regarded as that of a fictitious particle moving over the wave surface following the instantaneous maximum of kinetic energy.

The \emph{apparent horizon} is defined as the locus $r = r_{\mathrm{AH}}$ where light rays momentarily stop as observed by a distant observer:
\begin{equation}
C_e(r_{\mathrm{AH}}) - U(r_{\mathrm{AH}}) = 0.
\end{equation}
Intuitively, a spaceship located at $r_{\mathrm{AH}}$ sends a light pulse outward; the pulse does not propagate instantaneously outward but hovers at the horizon. Inside the region where $C_e < U$, no outward trajectory is possible.  This is the gravity analogue of the \emph{local breaking onset} of the wave. There is no singularity at $r = r_{\mathrm{AH}}$: the Ricci curvature remains finite and null rays cross the apparent horizon smoothly. The true singularity of the effective geometry occurs where the Ricci scalar diverges~(see, e.g.,~\cite{MisnerThorneWheeler1973,Carroll2004}).

The \emph{event horizon} is the global boundary separating light rays that ultimately escape from those that never do. 
It corresponds to the last null geodesic that reaches distant observers.  
Unlike the apparent horizon, its location depends on the full future evolution of the wave field~\citep{MisnerThorneWheeler1973,Thornburg2007EventHorizon}.

In gravitational collapse of black holes (the analogue of waves that inevitably break), the event horizon is typically crossed \emph{before} the apparent horizon. The event horizon is the global boundary beyond which escape is no longer possible, even though nothing catastrophic happens locally at that point. By contrast, the apparent horizon is \emph{local} where light rays momentarily stop~\citep{Thornburg2007EventHorizon}.
A key point is that a light pulse crossing the apparent horizon only \emph{locally} realizes that it can no longer escape, but the causal trap has already begun earlier: the pulse entered the black-hole region when it crossed the event horizon. 
Thus, the apparent horizon~(breaking onset) is where light~(wave) \emph{realizes too late} that escape is impossible, while the event horizon~(breaking inception) marks the true point of no return.

We do not pursue this gravity analogy further, which provides a geometrization of the fold catastrophe  of breaking waves and offers further physical insight into why and how waves break.

\section{Conclusions}
We have developed a new dynamical-systems framework for wave breaking in ideal, incompressible, and irrotational free-surface flows, grounded in first principles and supported by numerical simulations of unsteady wave groups. 

A central novelty of this work is the identification of a natural \emph{slow–fast structure} in the Euler dynamics when tracking the energetic hotspot on the wave surface. Near breaking, the surface slope~$m$ evolves on a fast timescale governed by the small parameter $\epsilon = 1/U_z$, the inverse of the horizontal velocity gradient, while the focusing parameter $A$ vary slowly, among others. 
This slow–fast system reveals a fold catastrophe in the $(m,A)$ parameter space, whose boundary provides the geometric skeleton of the dynamics near the breaking region. Finite-time blowup occurs as the boundary is crossed, marking the breaking onset. 

Extending the fast–slow model~\eqref{ODEdim4b} beyond the breaking region requires a multiple-scale perturbation method to capture higher-order surface-slope effects and to determine the hotspot speed~$C_e$ in Eq.~\eqref{Ce2} away from the turning point~$O$~(see Fig.~\ref{FIG3fold}). In this broader regime, the fold catastrophe may represent a cross-section of a cubic cusp or even a higher-order swallowtail catastrophe~\citep{Thom1975,zeeman1976catastrophe,arnold1984catastrophe}.

The inception of breaking is identified as associated with the crossing of a well-defined threshold at the characteristic slope angle $\theta^*=\arctan{\sqrt{2}-1}=22.5^{\circ}$. This critical angle marks the maximum anisotropy between the Hessians of the velocity and pressure fields that can be sustained by a wave. Physically, this anisotropy reflects an imbalance between the kinetic and potential energy fluxes that govern the local wave dynamics.  This trend is robust, as it consistently appears across other simulated breaking wave groups reported in~\citep{BarthelemyJFM2018}. A similar anisotropy mechanism underlies the emergence of the  $30^{\circ}$ slope threshold for breaking inception at the inflection point, beyond which the anisotropy increases. We point out that there is no catastrophe that guides the wave dynamics following the inflection point.

The crest height of a breaking wave is ultimately bounded by a maximum excess of kinetic energy over potential energy that the flow cannot sustain. It surpasses the classical Stokes limit $U^2/2g$ due to the additional local acceleration generated by the wave’s nonstationarity. Rogue waves, being strongly transient, may easily exceed this bound and are therefore inherently prone to breaking.



Wave breaking can be also interpreted as a gravity analogue of collapsing black holes, with the apparent and event horizons corresponding to the breaking onset and inception thresholds. This geometrization may provide further insight in the fold catastrophe of breaking waves and the resulting energy focusing on the wave surface.

In summary, the fold-catastrophe framework provides a unified geometric description of wave steepening and breaking. This perspective clarifies the bifurcation underlying wave breaking and lays the foundation for future investigations for three-dimensional waves, including the effects of wind, viscosity, and surface tension, ultimately improving predictive wave models and understanding rogue waves in realistic ocean conditions.

\section{Acknowledgments}
The author gratefully acknowledges X. Barthelemy and D. Boettger for generously sharing their numerical data on breaking and non-breaking wave groups. Special thanks are extended to D. Boettger and M. L. Banner for their support and insightful discussions during the preparation of this manuscript. The author thanks Cristel Chandre for revising an early draft of the manuscript and for suggesting the derivation of asymptotic dynamical models valid near breaking. 
The author also thanks D. Clamond and D. Dutykh for insightful discussions on steady Stokes waves and group velocities. A special thanks to Arash Yavari, Cristel Chandre and Chongchun Zeng for insightful discussions on geometry and dynamical systems. 

\section{Funding}
This research received no specific grant from any funding agency, commercial or not-for-profit sectors.

\section{Declaration of Competing Interests}
The author declares none.

\section{Data Availability Statement}
The data that support the findings of this study are available from X. Barthelemy and D. Boettger. 

\section*{Author Contribution Statement}
{\bf F. Fedele}: Conceptualization, Methodology, Software, Writing - Original Draft.

\appendix

\section{Time evolution of local observables at a moving surface point}\label{appA}

We study the surface dynamics by tracking the values of physical fields, or \emph{observables}, in time at a specific moving point $Q$ on the wave surface, of coordinates
\[
\mathbf{X}_Q(t) =\left(x = x_Q(t), \, z = \eta(x_Q(t),t)\right)\,.
\]
This defines a trajectory that is not Lagrangian, as it follows a geometrically distinguished point on the wave profile rather than a fluid particle.

The specific point~$Q$ is characterized by the generic condition 
\[
R[\eta, \mathbf{U}](x,t) = 0\,,
\]
where $R[\eta, \mathbf{U}]$ is a functional of the field variables and their derivatives. The associated velocity vector is 
$\mathbf{V}_Q = (C, \dot{z}_Q)$~\citep[see, e.g.,][]{FedeleEPL2014}, 
where the horizontal component is
\[
C = \frac{dx_Q}{dt} = \dot{x}_Q = -\frac{\partial_t R}{\partial_x R}\,, 
\qquad  
\dot{z}_Q = \partial_t \eta + C\,\partial_x \eta\,,
\]
with all derivatives evaluated at the moving point $Q$. For example, \cite{BarthelemyJFM2018} investigated breaking inception by tracking a wave crest where $R = \eta_x = 0$, with 
$C = -\partial_{xt} \eta / \partial_{xx} \eta$~\citep{FedeleEPL2014,FedeleJFMcrestspeed2020}.  
\cite{Seliger1968} and \cite{McAllister2023} tracked surface points of maximum inflection, 
where $R = \partial_{xx} \eta = 0$ and 
$C = -\eta_{xxt} / \eta_{xxx}$.  
In contrast, \citet{BoettgerJFM2023,BoettgerPRF2024} followed surface points of maximum kinetic energy. 

Consider a field observable~$f(x,z,t)$ tracked at the moving point~$Q$ on the surface. 
Its local value is 
\[
F(t) = f\left(x = x_Q(t), \, z = \eta(x_Q(t),t), \, t\right)\,.
\]
The time evolution of $F(t)$ depends on the variations of the field~$f$ on the surface at~$z=\eta$ 
and on the speed $C$:
\[
\frac{dF}{dt}  
=  \frac{\partial f}{\partial t} 
+ \frac{\partial f}{\partial x} C 
+ \frac{\partial f}{\partial z} 
\left( \frac{\partial \eta}{\partial x} C + \frac{\partial \eta}{\partial t} \right)\,,\qquad z=\eta(x_Q,t),
\]
where all partial derivatives are evaluated at the moving point~$Q$. Drawing on~\cite{Bridges_2009}, it is convenient to define the associated one-dimensional~($1$D) observable 
\[
\tilde{f}(x,t) = f\left(x, z = \eta(x,t), t\right)\,,
\]
measured on the surface. Then,
\begin{equation}\label{dF/dt}
\frac{dF}{dt}  =  \frac{\partial \tilde{f}}{\partial x}\, C  + \frac{\partial \tilde{f}}{\partial t}\,,
\end{equation}
where
\[
\frac{\partial \tilde{f}}{\partial x}  
=  \frac{\partial f}{\partial x}
+ \frac{\partial f}{\partial z} \frac{\partial \eta}{\partial x}\,,
\qquad  
\frac{\partial \tilde{f}}{\partial t}  
=  \frac{\partial f}{\partial t} 
+ \frac{\partial f}{\partial z} \frac{\partial \eta}{\partial t}\,,
\]
are the space and time derivatives of the one-dimensional field $\tilde{f}(x,t)$.  

The time derivative~\eqref{dF/dt} can be expressed in terms of Lagrangian derivatives. Since $\partial_t = D/Dt - \mathbf{u}\cdot\nabla$, it follows that
\[
\frac{dF}{dt}  
=  \frac{D f}{D t} - u\,\frac{\partial f}{\partial x} - w\,\frac{\partial f}{\partial z} 
+ \frac{\partial f}{\partial x}\, C 
+ \frac{\partial f}{\partial z} 
\left( \frac{\partial \eta}{\partial x}\, C + \frac{\partial \eta}{\partial t} \right)\,, 
\qquad z=\eta(x_Q,t)\,,
\]
or equivalently,
\[
\frac{dF}{dt}  
=  \frac{D f}{D t} - (u-C)\,\frac{\partial f}{\partial x}
+ \frac{\partial f}{\partial z} 
\left( \frac{\partial \eta}{\partial t} + \frac{\partial \eta}{\partial x}\, C - w \right)\,, 
\qquad z=\eta(x_Q,t)\,.
\]

Using the kinematic boundary condition~\eqref{kin}, this simplifies to
\begin{equation}\label{dF/dtLagrangian}
\frac{dF}{dt}  
=  \frac{D f}{D t} - (u-C)\left( \frac{\partial f}{\partial x} + \frac{\partial \eta}{\partial x}\,\frac{\partial f}{\partial z}\right)\,, 
\qquad z=\eta(x_Q,t)\,.
\end{equation}

Thus, the time derivative operator along the free surface is
\[
\frac{d}{dt} = \frac{D}{Dt} - (u-C)\left( \partial_x + \eta_x \,\partial_z \right)\,, 
\qquad z=\eta(x_Q,t)\,.
\]
In the following, either this expression or the form in~\eqref{dF/dt} is used conveniently for simplified derivations.


We consider the following local observables. The crest elevation~$h$, from the $z=0$, observed at the moving point~$Q$ and defined from the $1D$ field observable~$\eta$ as
\begin{equation}\label{ht}
   h(t) =  \eta(x_Q(t), t)\,. 
\end{equation}
The surface slope~$m$ is defined from the $1D$ field observable~$\eta_x$ as
\begin{equation}\label{mt}
    m(t) =  \eta_x(x_Q(t), t)\,.
\end{equation}
The local fluid velocities~$\mathbf{U}=(U,W)$ are given by
\begin{equation}\label{Ut}
   U(t) = \tilde{u}(x_Q(t),t)\,, \quad   W(t) = \tilde{w}(x_Q(t),t)\,,
\end{equation}
and follow from the corresponding $1D$ Eulerian fluid velocity fields on the surface
\begin{equation}\label{1DUWfields}
    \tilde{u}(x,t) = u(x, \eta(x,t),t)\,,\qquad \tilde{w}(x,t) = w(x, \eta(x,t),t)\,.
\end{equation}
We also consider the local gradient components of~$\nabla u= (u_x,u_z)$ tracked at $Q$ and given by
\begin{equation}\label{UxUz}
   U_x(t) = \tilde{u}_x(x_Q(t),t)\,, \quad   U_z(t) = \tilde{u}_z(x_Q(t),t)\,,
\end{equation}
which follow from the corresponding $1D$ fields on the surface
\begin{equation}\label{1DUxUzfields}
    \tilde{u}_x(x,t) = u_x(x, \eta(x,t),t)\,,\qquad \tilde{u}_z(x,t) = u_z(x, \eta(x,t),t)\,.
\end{equation}
The time evolution of each of these local observables depend on the others due the nonlinear nature of the wave dynamics of the Euler equations. 

\subsection{Surface slope~$m$}\label{appAm}
From Eq.~\eqref{mt}, the rate of change of the surface slope~$m$ follows from time differentiation as
\[
\frac{dm}{dt} = (C\, \partial_{xx} + \partial_{xt})\eta\,,\quad \text{at}\,\,x=x_Q(t)\,,
\]
where the partial derivatives are evaluated at the tracking point~$Q$. The mixed derivative follows by space-differentiating the kinematic condition~\eqref{kin} as
\[
  \eta_{xt} = \partial_x\big[(- u\, \eta_x +  w)\big|_{z=\eta}\big]\,,
\]
from which 
\[
  \eta_{xt} = -u \eta_{xx} -\eta_x (u_x + u_z \eta_x) +  (w_x + w_z \eta_x)\,,\quad \text{at}\,\,z=\eta\,.
\]
Imposing mass conservation~$w_z = -u_x$ and irrotationality~$w_x = u_z$ yield
\begin{align}\label{etaxt}
\eta_{xt} & = -u \eta_{xx} -\eta_x (u_x + u_z \eta_x) +  (u_z - u_x \eta_x)\\
& =  -u \eta_{xx} -2\, u_x\eta_x +u_z(1 - \eta_x^2)\,,\qquad\qquad \text{at}\,\,z=\eta\,. \nonumber
\end{align}
Thus,
\begin{equation}\label{mslope2}
    \frac{dm}{dt} = -2\, U_x\eta_x +U_z(1 - \eta_x^2) - (U-C)\, \eta_{xx}|_Q\,, 
\end{equation}

Note that~$\eta_{xx}=\kappa (1+\eta_x^2)^{3/2}$ where the extrinsic surface curvature $\kappa = \eta_{xx} /(1+ \eta_x^2)^{3/2}$. The slope equation~\eqref{mslope2} is not closed as its evolution depends on other observables. 

\subsection{Crest height~$h$}\label{appAh}
From Eq.~\eqref{ht} 
\[
\frac{dh}{dt} = C\, \partial_{x}\eta + \partial_{t}\eta\,.
\]
From the kinematic condition~\eqref{kin}
\[
  \eta_{t} = - u\, \eta_x +  w\,,\qquad \text{at}\,\,z=\eta\,,
\]
from which
\begin{equation}\label{crestheight}
    \frac{dh}{dt} = -(U-C)\, m + W\,.
\end{equation}

\subsection{Fluid velocities~$U$ and $W$}\label{appAUW}
From~\eqref{Ut}, 
\begin{equation}
\frac{dU}{dt} = C\, \partial_x \tilde{u}  + \partial_t \tilde{u}\,,
\end{equation}
where the partial derivatives, evaluated at the tracking point~$Q$, are given by
\[
\partial_x \tilde{u} = u_x + u_z\, \eta_x\,,\qquad \text{at}\,\,z=\eta\,, 
\]
and 
\begin{align*}
\partial_t \tilde{u} &= u_t + u_z\,\eta_t\, \\
& =  - u u_x - w u_z  - (\tilde{p}/\rho)_x  + u_z ( - u\,\eta_x + w  )\\ 
& = - u u_x  - u u_z\,\eta_x  - (\tilde{p}/\rho)_x,\,\qquad\qquad\qquad\qquad   \text{at}\,\,z=\eta\,,
\end{align*}
where we have used the momentum equations~\eqref{Euler}. The rate of change over time of~$U$ follows as 
\begin{equation}\label{dUdt}
\frac{dU}{dt} =  - (U-C) (U_x + m U_z)  - (\tilde{p}_x/\rho)\big|_Q\,.
\end{equation}
Similarly,
\[
\frac{dW}{dt} =  - (U-C) (W_x + m W_z)  - (\tilde{p}_z/\rho)\big|_Q-g\,,
\]
Using the relations~$w_z = -u_x$ and~$w_x = u_z$ from mass conservation and irrotationality,  
\begin{equation}\label{dWdt}
\frac{dW}{dt} =  - (U-C) (U_z - m U_x)  - (\tilde{p}_z/\rho)\big|_Q\,,
\end{equation}
where the pressure gradients are evaluated at~$Q$. 

A direct evaluation of these expressions follows by applying the total time derivative in Eq.~\eqref{dF/dtLagrangian} together with the momentum equations~\eqref{Euler}:
\begin{align}\label{dU/dtLagrangian}
\frac{d\mathbf{U}}{dt}  
&=  \frac{D\mathbf{u}}{Dt} - (u-C)\left( \frac{\partial \mathbf{u}}{\partial x} + \frac{\partial \eta}{\partial x}\,\frac{\partial \mathbf{u}}{\partial z}\right) \\
&= -\nabla(p/\rho+gz) - (u-C)\left( \frac{\partial \mathbf{u}}{\partial x} + \frac{\partial \eta}{\partial x}\,\frac{\partial \mathbf{u}}{\partial z}\right)\,, 
\qquad z=\eta(x_Q,t)\,.
\end{align}

At the moving point~$Q$ on the free surface, where $m=\eta_x$, we have $\tilde{\mathbf{u}}\big|_Q=\mathbf{U}=(U,W)$, i.e. the Eulerian velocity field evaluated along the trajectory of $Q$. The corresponding velocity gradients are
\[
 \frac{\partial \mathbf{u}}{\partial x}\Big|_Q =(U_x,W_x)=(U_x,U_z)\,, 
 \qquad  
 \frac{\partial \mathbf{u}}{\partial z}\Big|_Q=(U_z,W_z)=(U_z,-U_x)\,,
\]
where the last equalities follow from incompressibility and irrotationality. 
Substituting these relations into Eq.~\eqref{dU/dtLagrangian} gives the time evolution of the Eulerian velocity components tracked at~$Q$.

\subsection{Kinetic energy density $K_e$}\label{appAKe}
The kinetic energy density for unit mass tracked at a moving point $Q$ is defined from the $1D$ field observable
\begin{equation}\label{1dkefield}
    \tilde{k}_e(x,t)=k_e(x,z=\eta(x,t),t)\,,
\end{equation}
where $k_e=(u^2+w^2)/2$ is the kinetic energy density of the fluid. Thus, 
\[
    K_e(t) = \tilde{k}_e(x=x_Q(t),t)=(U(t)^2 + W(t)^2)/2\,,
\]
where~$\mathbf{U}=(U,W)$ are the Eulerian velocity components tracked at $Q$. Then,
\[
\frac{dK_e}{dt} = U \frac{dU}{dt} + W \frac{dW}{dt}\,,
\]
and from Eqs.~\eqref{dUdt} and~\eqref{dWdt},
\begin{equation}\label{dKedt}
    \frac{dK_e}{dt} = - (U-C) [U_x ( U - W\,m )  + U_z ( W + U\,m )]  - \mathbf{U}\cdot \nabla(p/\rho)\big|_Q\,.
\end{equation}
Alternatively, from~\eqref{1dkefield},
\begin{equation}
\frac{dK_e}{dt} = C\frac{\partial \tilde{k}_e}{\partial x}  + \frac{\partial \tilde{k}_e}{\partial t}\,,
\end{equation}
where the partial derivatives are evaluated at the tracking point~$Q$. We have
\begin{align*}
\frac{\partial \tilde{k_e}}{\partial x}
&= \left.\frac{\partial k_e}{\partial x}\right|_{z=\eta} + \left.\frac{\partial k_e}{\partial z}\right|_{z=\eta} \frac{\partial \eta}{\partial x}\\
&=  \left.(u u_x+ w w_x)\right|_{z=\eta} 
  + \left. (u u_z + w w_z)\right|_{z=\eta} \,\eta_x \\
&= \left.(u u_x+ w u_z)\right|_{z=\eta} 
  + \left. (u u_z - w u_x)\right|_{z=\eta} \,\eta_x\\
&=u_x ( u - w\,\eta_x )  + u_z ( w + u\, \eta_x )\,,\qquad\qquad \text{at}\,\,z=\eta\,,
\end{align*}
and 
\begin{align*}
 \frac{\partial \tilde{k_e}}{\partial t}
 &= \left.\frac{\partial k_e}{\partial t}\right|_{z=\eta} + \left.\frac{\partial k_e}{\partial z}\right|_{z=\eta} \frac{\partial \eta}{\partial t} \\
 &= \left. (u u_t + w w_t)\right|_{z=\eta}  + \left. (u u_z + w w_z)\right|_{z=\eta}\, \eta_t \\
 &= \left. (u u_t + w w_t)\right|_{z=\eta}  + \left. (u u_z -  w u_x)\right|_{z=\eta}\, \eta_t\,,\qquad \text{at}\,\,z=\eta\,.
\end{align*}
Here we have used~$w_z = -u_x$ and~$w_x = u_z$ from mass conservation and irrotationality. 
Using the momentum equations~\eqref{Euler} the two terms 
\begin{align*}
u u_t + w w_t & = u ( - u u_x - w u_z  - (p/\rho)_x ) + w ( - u w_x - w w_z  - (p/\rho + g z)_z )\\
& = u ( - u u_x - w u_z  - (p/\rho)_x ) + w ( - u u_z + w u_x  - (p/\rho + g z)_z )\\
&=u_x (-u^2 + w^2) - 2 u_z  u w  - \mathbf{u}\cdot \nabla(p/\rho + g z)\,,\qquad\qquad\qquad\qquad \text{at}\,\,z=\eta\,,
\end{align*}
and
\begin{align*}
(u u_z -  w u_x)\, \eta_t & = (u u_z -  w u_x) (- u \eta_x + w)\\ 
& = w (u u_z -  w u_x) - u \eta_x (u u_z -  w u_x)\\
& = u_x (-w^2 + u w\, \eta_x) + u_z (u w - u^2 \eta_x)\,,\qquad \text{at}\,\,z=\eta
\end{align*}
can be simplified. Then,
\begin{align*}
 \frac{\partial \tilde{k_e}}{\partial t}
 &= \left. (u u_t + w w_t)\right|_{z=\eta}  + \left. (u u_z -  w u_x)\right|_{z=\eta}\, \eta_t\\
 &=u_x (-w^2 + u w\, \eta_x) + u_z (u w - u^2 \eta_x) +  u_x (-u^2 + w^2) - 2 u_z  u w  - \mathbf{u}\cdot \nabla(p/\rho + g z)\\
 & = u_x ( u w\, \eta_x) + u_z ( - u^2 \eta_x) +  u_x (-u^2) -  u_z  u w  - \mathbf{u}\cdot \nabla(p/\rho + g z)\\
 & = u_x ( u w\, \eta_x - u^2) + u_z ( - u^2 \eta_x - u w)   - \mathbf{u}\cdot \nabla(p/\rho + g z)\\
 & = u u_x ( w\, \eta_x - u) + u u_z ( - u \eta_x - w)   - \mathbf{u}\cdot \nabla(p/\rho + g z)\\
 &=  - u [  u_x ( u - w\, \eta_x ) + u_z ( w +  u \eta_x )]   - \mathbf{u}\cdot \nabla(p/\rho + g z)\\
 & = - u \frac{\partial \tilde{k_e}}{\partial x} - \mathbf{u}\cdot \nabla(p/\rho + g z)\,,\qquad \text{at}\,\,z=\eta\,,\\
\end{align*}
from which
\[
\frac{\partial \tilde{k_e}}{\partial t} = - u \frac{\partial \tilde{k_e}}{\partial x} - [\mathbf{u}\cdot \nabla(p/\rho + g z)]|_{z=\eta}\,.
\]

At the moving point~$Q$ on the free surface, where $m=\eta_x$, we have
\begin{equation}\label{UandgradU_at_Q}
    \mathbf{u}|_Q=\mathbf{U}=(U,W), 
\qquad
\nabla u|_Q=(\partial_x u,\partial_z u)\big|_Q=(U_x,U_z)\,,
\end{equation}
and 
\begin{equation}\label{grafP_at_Q}
    \nabla p\big|_Q=(\partial_x p,\partial_z p)\big|_Q=(P_x,P_z)\,,
\end{equation}
evaluated along the trajectory of $Q$. Then, the rate of change over time of~$K_e$ follows as 
\begin{equation}\label{Ke2}
\frac{dK_e}{dt} = C\frac{\partial \tilde{k}_e}{\partial x}  + \frac{\partial \tilde{k}_e}{\partial t} = -(U -C) \left. \frac{\partial \tilde{k}_e}{\partial x}\right|_{x=x_Q(t)}- \mathbf{U}\cdot \nabla(p/\rho + g z)\big|_Q\,,
\end{equation}
where $\nabla(p/\rho + g z)\big|_Q=(P_x/\rho, P_z/\rho + g)$ and the spatial kinetic energy gradient at~$Q$ 
\begin{equation}\label{KEspacegradient}
\left. \frac{\partial \tilde{k}_e}{\partial x}\right|_{x=x_Q(t)} 
    =  U_x \,( U - W\,m )  + U_z \,( W + U\,m )\,.
\end{equation}

A direct evaluation of these expressions follows by applying the semi-Lagrangian derivative in Eq.~\eqref{dF/dtLagrangian} together with the momentum equations~\eqref{Euler}:
\begin{align}\label{dKe/dtLagrangian}
\frac{d K_e}{dt}  
&=  \frac{D k_e}{Dt} - (u-C)\left( \frac{\partial k_e}{\partial x} + \frac{\partial \eta}{\partial x}\,\frac{\partial k_e}{\partial z}\right) \\
&= -\mathbf{u}\cdot\nabla(p/\rho+gz) - (u-C)\left( u u_x + w w_x + \eta_x (u u_z + w w_z)\right) \\
&= -\mathbf{u}\cdot\nabla(p/\rho+gz) - (u-C)\left( u u_x + w u_z + \eta_x (u u_z - w u_x)\right) \\
&= -\mathbf{u}\cdot\nabla(p/\rho+gz) - (u-C)\left( u_x(u-\eta_x w) + u_z (w + \eta_x u)\right) \\
&= -\mathbf{u}\cdot\nabla(p/\rho+gz) - (u-C)\frac{\partial \tilde{k_e}}{\partial x}\,,
\qquad z=\eta(x_Q,t)\,,
\end{align}
where we have used incompressibility and irrotationality constraints. Substituting the relations in Eqs.~\eqref{UandgradU_at_Q} and~\eqref{grafP_at_Q} into Eq.~\eqref{dKe/dtLagrangian} gives the time evolution of the kinetic energy tracked at~$Q$ in Eq.~\eqref{Ke2}.

\subsection{Velocity gradients $U_x,U_z$}\label{appAUxUz}

For inviscid fluid flows, the Lagrangian derivative of the velocity gradient tensor~$\nabla \mathbf{u}=[\partial_a u_b]$ is~\citep{ohkitani1995pressureHessian,Chellard2006pressureHessian}
\begin{equation}
\frac{D\nabla \mathbf{u}}{Dt} = - (\nabla \mathbf{u})^\top  \nabla \mathbf{u}-\frac{1}{\rho} \mathcal{H}(p)\,,
\label{eq:grad_u_euler}
\end{equation}
where $\mathbf{u} = (u, w)$ is the two-dimensional velocity field in the $(x,z)$-plane, $D/Dt = \partial_t + \mathbf{u}\cdot \nabla$, $^\top$ denotes matrix transpose and the pressure Hessian~$\mathcal{H}(p)=[\partial_{ab}p]$. 
These equations describe how the strain rates along the coordinate directions evolve under the influence of the competing pressure gradients and self-interaction of velocity gradients~(nonlinear advection), a hallmark of vortex stretching and deformation in inviscid flows.

For potential flows, we just need the the Lagrangian derivatives of the gradient $\nabla u=(\partial_x u, \partial_z u)$. The other two components~($\partial_x w=\partial_z u, \partial_z w= -\partial_x u$) of the velocity gradient tensor follow from the irrotationality and incompressibility conditions.  From \eqref{eq:grad_u_euler},
\begin{equation}
\frac{Du_x}{Dt}= - ( u_x^2 + u_z^2)-\frac{1}{\rho} p_{xx}\,,
\label{eq:Dt_duxdx}
\end{equation}
and
\begin{equation}
\frac{D u_z}{Dt} =  -\frac{1}{\rho} p_{xz}\,.
\label{eq:Dt_duzdz}
\end{equation}

Applying the total time derivative in Eq.~\eqref{dF/dtLagrangian} yields
\begin{align}\label{d_Ux/dtLagrangian}
\frac{d U_x}{dt}  
&=  \frac{Du_x}{Dt} - (u-C)\left( \frac{\partial u_x}{\partial x} + \frac{\partial \eta}{\partial x}\,\frac{\partial u_x}{\partial z}\right) \\
&= - ( u_x^2 + u_z^2)-\frac{1}{\rho} p_{xx} - (u-C)\left( u_{xx} + \eta_x u_{xz}\right)\,,
\qquad z=\eta(x_Q,t)\,,
\end{align}
where the mixed derivative~$u_{xz}=u_{zx}$. 

Similarly,
\begin{align}\label{d_Uz/dtLagrangian}
\frac{d U_z}{dt}  
&=  \frac{Du _z}{Dt} - (u-C)\left( \frac{\partial u_z}{\partial x} + \frac{\partial \eta}{\partial x}\,\frac{\partial u_z}{\partial z}\right) \\
&= -\frac{1}{\rho} p_{xz} - (u-C)\left( u_{xz} + \eta_x u_{zz}\right)\,,
\qquad z=\eta(x_Q,t)\,.
\end{align}

At the moving point~$Q$ on the free surface, where $m=\eta_x$, 
\begin{equation}\label{UxUzODE}
\left\{
\begin{aligned}
\frac{dU_x}{dt} &=  - (U-C)(U_{xx} + U_{xz} m) -(U_x^2 + U_z^2) - (p_{xx}/\rho)\big|_Q\,,\\
\frac{dU_z}{dt} &=  - (U-C)(U_{xz} + U_{zz} m)  - (p_{xz}/\rho)\big|_Q\,.\\
\end{aligned}
\right.
\end{equation}
 Using the Poisson equation~\eqref{eq:PoissonP2}, we can express the velocity gradients  in terms of pressure derivatives as
\begin{equation}\label{dUxdt2}
 \frac{dU_x}{dt} =  - (U-C)(U_{xx} + U_{xz} m) +\frac{-P_{xx}+P_{zz}}{2\rho}\,.  
\end{equation}

Note that the evolution of velocity gradients depends on the higher order gradients of pressure and fluid velocities~$ (u_{xx},u_{xz},u_{zz})|_Q=(U_{xx},U_{xz},U_{zz})$ and $(p_{xx},p_{xz},p_{zz)}|_Q=(P_{xx},P_{xz},P_{zz})$ at~$Q$. This is due to the nonlinear nature of the Euler equations and the associated closure problem. 

\subsection{Velocity $C_e$ of the energetic hotspot}\label{C_e_derivation}
At the energetic hotspot, the maximum kinetic energy is attained, where the spatial kinetic energy gradient~\eqref{KEspacegradient}  vanishes:
\begin{equation}\label{KEspacegradient2}
\partial_x \tilde{k}_e\big|_Q
    =  U_x \,( U - W\,m )  + U_z \,( W + U\,m )=0\,.
\end{equation}
The energetic hotspot speed ~$C_e=C=dx_Q/dt$ follows from
\begin{align*}\label{KExdt}
\frac{d\,\partial_x \tilde{k}_e\big|_Q}{dt} &=
\frac{dU_x}{dt}( U - W\,m ) + \frac{dU_z}{dt}( W + U\,m ) +\\
&+\frac{dU}{dt}( U_x + m U_z ) + \frac{dW}{dt}( U_z - m U_x ) +\\
&  +\frac{dm}{dt}( U U_z - W U_x )=0\,.
\end{align*}
Substituting the time derivatives from Eqs.~\eqref{mslope2},\eqref{dUdt},\eqref{dWdt},\eqref{UxUzODE},\eqref{dUxdt2} and $W=s U$,$U_x= m_2 U_z$, we get 
\begin{equation}\label{Ce2}
C_e = U + \frac{ \mathrm{trace}(G_p\,\mathcal{H}(p)\big|_Q)/\rho+\frac{(1 + m^2) ( g + (1 + m s ) P_z/\rho )}{U(1 - m s)}U_z -\frac{ (1+m^2) (1+m s) (s^2+1) }{(1- m s)^2}U_z^2 }{\mathrm{trace}(G_u\,\mathcal{H}(u)\big|_Q) +\frac{(s^2+1)\eta_{xx}|_Q  }{1- m s}U_z+\frac{(m^2+1)^2 (s^2+1) }{U(1- m s)^2}U_z^2 }\,,
\end{equation}
where
\[
\mathrm{trace}(G_f\,\mathcal{H}(f))=\sum_{ab} g_f^{ab} H_{ab}(f)\,,
\]
the Hessian matrices tracked at the energetic hotspot~$Q$
\begin{equation}\label{Hessians}
 \mathcal{H}(u)\big|_Q=[ H^{(u)}_{ab}]=[\partial_{ab} u|_Q] = \begin{pmatrix}U_{xx}&U_{zx}\\[2mm] U_{zx}&U_{zz}\end{pmatrix}\,,\qquad \mathcal{H}(p)\big|_Q=[ H^{(p)}_{ab}]=[\partial_{ab} p|_Q] = \begin{pmatrix}P_{xx}&P_{zx}\\[2mm] P_{zx}&P_{zz}\end{pmatrix}\,,   
\end{equation}
and the symmetric $2\times 2$ tensors
\[
G_u=[g_u^{ab}] = \begin{pmatrix}g^{xx}_u & g_u^{xz}\\[1mm] g_u^{zx} & g_u^{zz}\end{pmatrix} = 
\begin{pmatrix}1-ms & \frac{1}{2}(s+2m-m^2s)\\[1mm] \frac{1}{2}(s+2m-m^2s) & m(m+s)\end{pmatrix}\,,
\]
with determinant~$g_u=\det{G_u}= -s^2(1 +m^2)^2/4$,
and 
\[
G_p=[g_p^{ab}] = \begin{pmatrix}g^{xx}_p & g_p^{xz}\\[1mm] g_p^{zx} & g_p^{zz}\end{pmatrix} =\frac{1}{2} 
\begin{pmatrix} 1-ms & m+s\\[1mm] m+s & m s -1\end{pmatrix}\,,
\]
with determinant~$g_p=-(1+m^2)(1+s^2)/4$.
Their inverses
\[
G_u^{-1} = [g^{(u)}_{ab}] = \frac{1}{g_u} \begin{pmatrix} m(m+s) & -\frac{1}{2}(s+2m-m^2s)\\[1mm] -\frac{1}{2}(s+2m-m^2s) & 1-ms\end{pmatrix}\,,
\]
 and 
\[
G_p^{-1} = [g^{(p)}_{ab}] = \frac{2}{g_p} 
\begin{pmatrix} -1 + ms & -m-s\\[1mm] -m-s & 1 - m s\end{pmatrix}\,,
\]
are the matrix representations of the following two metrics in coordinates $(x,z)$:
\[
ds_u^2 = 4\frac{- m (m+s)\, dx^2 +(m s-1)\, dz^2  + (2 m+s(1-m^2))\, dx\, dz}{\left(m^2+1\right)^2 s^2}\,,
\]
and
\[
ds_p^2 = 2\frac{(1 - m s)(dx^2 - dz^2) + 2 (m+s) dx\, dz}{(1+m^2)(1+s^2)}\,.
\]
These metrics are pseudo-Riemannian, specifically \emph{hyperbolic} where their determinant does not vanish. They encode the anisotropy of the wave surface dynamics to the underlying velocity and pressure fields.

Note that the trace of the velocity Hessian is null, because $u$ is an harmonic function. Moreover, the pressure Hessian may be simplified using the pressure relation in Eq.~\eqref{Pressurerelation2} and the Poisson equation~\eqref{eq:PoissonP2}. This will not be pursued here.

\subsection{Surface curvature~$\eta_{xx}|_Q$}\label{appetaxx}

The rate of change of the surface slope~$m$ follows from time differentiation as
\begin{equation}\label{detaxxQdt}
    \frac{d\,\eta_{xx}|_Q}{dt} = C\, \partial_{xxx}\eta + \partial_{xxt}\eta\,,\qquad \text{at}\,\,x=x_Q(t)\,,
\end{equation}
where the partial derivatives are evaluated at the tracking point~$Q$. 

The mixed derivative can be obtained by differentiating the kinematic condition~\eqref{etaxt} with respect to space:
\[
  \eta_{xxt} = \partial_x \left(-u \eta_{xx} - 2\, u_x \eta_x + u_z (1 - \eta_x^2)\right)\,,\qquad \text{at}\,\,z=\eta\,,
\]
from which
\begin{align}\label{etaxxt}
\eta_{xxt} &= -u\, \eta_{xxx} - \eta_{xx} (u_x + m u_z) - 2 \eta_{xx} (u_x + m u_z) \notag\\
&\quad + (1-m^2) (u_{zx} + m u_{zz}) - 2 m (u_{xx} + m u_{xz})\,, 
\quad \text{at } z=\eta\,,
\end{align}
where $u_{xz} = u_{zx}$. Thus, from Eq.~\eqref{detaxxQdt} one finds
\begin{equation}\label{etaxxQdt}
    \frac{d\,\eta_{xx}|_Q}{dt} = -\eta_{xxx} (U-C) - 3 \eta_{xx} (U_x + m U_z) - 2 m\, U_{xx} + m(1-m^2) U_{zz} + (1-3 m^2) U_{zx}\,,
\end{equation}
where 
\begin{equation}\label{UxxUzzUzx}
   U_{ab}(t) = \tilde{u}_{ab}(x_Q(t),t)\,,
\end{equation}
are the second-order derivatives of the horizontal fluid velocity tracked at $Q$. These derivatives are obtained from the corresponding 1D fields on the surface:
\begin{equation}\label{1DUxxUzzUzxfields}
    \tilde{u}_{ab}(x,t) = u_{ab}(x, \eta(x,t),t)\,.
\end{equation}

Due to the nonlinear nature of the Euler equations, the time evolution of these local observables depends on higher-order derivatives of the velocity and pressure fields.

We can rewrite~Eq.~\ref{etaxxQdt} as
\begin{equation}\label{etaxxQdt2}
    \frac{d\,\eta_{xx}|_Q}{dt} = \mathrm{trace}(G\,\mathcal{H}(u)\big|_Q) - \eta_{xxx}|_Q (U-C) - 3 \eta_{xx}|_Q (U_x + m U_z)\,, 
\end{equation}
where the trace
\[
\mathrm{trace}(G\,\mathcal{H}(u))=\sum_{ab} g^{ab} H_{ab}\,,
\]
the Hessian matrix at~$Q$
\[
\mathcal{H}(u)\big|_Q=[ H_{ab}]=[\partial_{ab} u|_Q] = \begin{pmatrix}U_{xx}&U_{zx}\\[2mm] U_{zx}&U_{zz}\end{pmatrix}\,,
\]
and the symmetric $2\times 2$ tensor $g^{ab}$ 
\[
G=[g^{ab}] = \begin{pmatrix}g_{xx}&g_{zx}\\[1mm] g_{zx}&g_{zz}\end{pmatrix} = 
\begin{pmatrix}-2m & \frac{1}{2}(1-3m^2)\\[1mm] \frac{1}{2}(1-3m^2) & m(1-m^2)\end{pmatrix}\,,
\]
with non-vanishing determinant~$g = -(1 + m^2)^2/4$. 
Its inverse
\[
G^{-1} = [g_{ab}] = \frac{1}{g} \begin{pmatrix} m(1-m^2) & -\frac{1}{2}(1-3m^2)\\[1mm] -\frac{1}{2}(1-3m^2) & -2m\end{pmatrix}\,,
\]
is the matrix representation of the metric in coordinates $(x,z)$:
\[
ds^2 = -\frac{4\, m(1-m^2)}{(1 +m^2)^2} dx^2 + \frac{8 m}{(1 +m^2)^2} dz^2 + \frac{4(1-m^3)}{(1 +m^2)^2} dx\, dz\,.
\]
This metric is pseudo-Riemannian, specifically \emph{hyperbolic}. It encodes the anisotropy of the surface response to the underlying velocity field. 

\subsection{Horizontal speed $C_{ip}$ of the inflection point}\label{appAnisotriopy}
If $Q$ is chosen as the inflection point, where $\eta_{xx}|_Q=0$, its horizontal speed $C_{ip}=C=dx_Q/dt$ satisfies
\[
\frac{d\,\eta_{xx}|_Q}{dt}=0\,.
\]
From Eq.~\eqref{etaxxQdt2}, 
\begin{equation}\label{Cipapp}
C_{ip}  = U +\frac{\mathrm{trace}(G\,\mathcal{H}(u)\big|_Q)}{\eta_{xxx}|_Q}
= U + \frac{-2 m\, U_{xx} + m(1-m^2) U_{zz} + (1-3 m^2) U_{zx}}{\eta_{xxx}|_Q}\,.
\end{equation}
Here the metric $g^{ab}$ acts as an anisotropic weight on the velocity Hessian, selecting the combinations of $U_{xx}, U_{zz}, U_{zx}$ that control the evolution of the inflection point speed and associated breaking inception.

\section{The governing ODE system}\label{appAODE}
For a generic moving point~$Q$, the time evolution of the abovementioned observables~$(m,h,U,W,U_x,U_z)$ satisfy the following set of ordinary differential equations~(ODEs):\footnote{In differential geometry, the collection of observables at the moving point $Q$ corresponds to a \emph{jet} at that point. The jet encapsulates the local behavior of the surface~$\eta$, the velocity fields~$U, W$, and their derivatives up to the desired order. The time evolution of the observables is determined by the evolution of these derivatives. This evolving jet captures the local geometry and dynamics around the moving point and governs the evolution of the observables.}
\begin{equation}\label{ODE2}
\left\{
\begin{aligned}
\frac{dm}{dt} &= -2\, U_x m + U_z(1 - m^2) - (U - C)\, \eta_{xx}|_Q \,, \\[4pt]
\frac{dh}{dt} &= - (U-C) m + W \,,\\[4pt]
\frac{dU}{dt} &=  - (U-C) (U_x + m U_z)  - (p/\rho)_x\big|_Q,\,\\[4pt]
\frac{dW}{dt} &=  - (U-C) (U_z - m U_x)  - (p/\rho + g z)_z\big|_Q\,,\\[4pt]
\frac{dU_x}{dt} &=  - (U-C)(U_{xx} + U_{xz} m) -(U_x^2 + U_z^2) - (p/\rho)_{xx}\big|_Q\,,\\[4pt]
\frac{dU_z}{dt} &=  - (U-C)(U_{zx} + U_{zz} m)  - (p/\rho)_{xz}\big|_Q\,,\\[4pt]
\cdots\\
\cdots\\
\end{aligned}
\right.
\end{equation}

The kinetic energy density depends on $U,W$ and it evolves in time according to
\begin{equation}\label{dKedt}
   \frac{dK_e}{dt} =
 -(U - C)[ U_x ( U - W\,m )  + U_z ( W + U\,m )]
- \mathbf{U} \cdot \nabla(\tilde{p}/\rho)\big|_Q \,, 
\end{equation}
where the term between squares in parentheses is the space energy gradient $\partial_x\tilde{k}_e|_{x=x_Q(t)}$ derived in Eq.~\eqref{KEspacegradient}.

Note that an additional relation between the two pressure gradient terms follows from differentiation  of the pressure dynamical condition~\eqref{Pdyn} along the surface tangent direction.

The system~\eqref{ODE2} is not closed as it depends also on the fluid velocity and pressure gradients, whose dynamical equations have non-trivial expressions and are omitted. This is the unpleasant side effect of the nonlinearities of the Euler equations. However, even including these equations closure will not be attained as those parameters depend on higher derivatives of fluid velocities and pressure.

\section{The slow-fast ODE system}\label{appAslowfastODE}
From Eq.~\eqref{timescaletau}, the rate of change of a generic observable $f(t)$ in the fast timescale~$\tau(t)=\int^t_{t_0} U_z(t')\, \text{d}\,t'$ is given by 
\[
\frac{df}{dt} =\frac{df}{d\tau}\frac{d\tau}{dt}=\frac{df}{d\tau}\frac{1}{\epsilon} \,,\qquad \epsilon\ll1\,,
\]
where $\epsilon=1/U_z\ll1$ near breaking. Then, the ODE system in Eq.~\eqref{ODE2} can be rewritten in terms of the dimensionless parameters~$s,m_2,A$ in Eqs.\eqref{s&m2}~and~\eqref{A} as
\begin{equation}\label{ODEdim}
\left\{
\begin{aligned}
\frac{dm}{d\tau} &= -2\, m_2 m + 1 - m^2 - A \,, \\[4pt]
\frac{dh}{d\tau} &= \epsilon \beta U ( - m + s ) \,,\\[4pt]
\frac{dU}{d\tau} &=   \beta U(1+ m\, m_2)  - \epsilon\, P_x/\rho\,,\\[4pt]
\frac{dW}{d\tau} &=   \beta U (1 - m\, m_2)  - \epsilon\,(P_z/\rho + g)\,,\\[4pt]
\frac{dU_x}{d\tau} &=   \epsilon \beta U (U_{xx} + U_{xz} m) - \epsilon^{-1}(1+m_2^2) - \epsilon P_{xx}/\rho\,,\\[4pt]
\frac{dU_z}{d\tau} &=  \epsilon U \beta(U_{zx} + U_{zz} m)  - \epsilon P_{xz}/\rho\,,\\[4pt]
\frac{ds}{d\tau} &=  s \left( \frac{1}{W} \frac{dW}{d\tau} - \frac{1}{U} \frac{dU}{d\tau}\right) \,,\\[4pt]
\frac{dm_2}{d\tau} &=  m_2 \left( \frac{1}{U_x} \frac{dU_x}{d\tau} - \frac{1}{U_z} \frac{dU_z}{d\tau}\right) \,,\\[4pt]
\cdots\\
\cdots
\end{aligned}
\right.
\end{equation}
where $\beta = (C-U)/U$ measures the relative fluid speed~(see Eq.~\eqref{beta} and section~$\S$~\ref{Intrinsictimescale}). 
The kinetic energy density evolves in time according to
\begin{equation}\label{dKedt}
   \frac{dK_e}{d\tau} =
  U^2( 1-\beta) \beta\,[ m_2 ( 1 - s\,m )  + ( s + \,m )]
- \epsilon\, \mathbf{U} \cdot \big[\nabla(p/\rho + g z)\big]\big|_Q \,. 
\end{equation}

The system~\eqref{ODEdim} is still not closed as these equations depend on gradients of fluid velocities and pressure. 
However, by tracking the wave dynamics at the energetic hotspot, the evolution of the surface slope~$m$ decouples from that of the momentum parameter $m_2$. Then, there is a clear slow-fast timescale separation, where geometric parameters evolve on the fast timescale~$\tau$, while kinematic parameters change adiabatically slow, as shown in the following section. 


\subsection{Tracking the energetic hotspot}\label{appB_hotspot}
Consider tracking the most energetic point~$Q$ on the wave surface, or hotspot, where the kinetic energy density $\tilde{k}_e$ is maximum~\citep{BoettgerJFM2023}. 
From Eq.~\eqref{KEspacegradient}, the energy gradient~$\partial_x \tilde{k}_e$ is null at the hotspot and defines the momentum parameter in~Eq.~\eqref{s&m2} as
\begin{equation}\label{m2eq2}
 m_2|_Q = \frac{U_x}{U_z} = \frac{U\,m + W}{W\,m - U} = \frac{m + s}{m\,s - 1}\,,
\end{equation}
as a function of solely the kinematic parameter $s=W/U$ and the surface slope~$m$. 

The speed $C_e=C=dx_Q/dt$ of the energetic hotspot follows from the condition
\begin{equation}\label{C_EP}
   \frac{d}{dt} \partial_x \tilde{k}_e \big(x_Q(t), t\big) = 0 \quad \Longrightarrow \quad 
   C_e = -\frac{\partial_{xt} \tilde{k}_e}{\partial_{xx} \tilde{k}_e} \,.
\end{equation}


The ODE system~\eqref{ODEdim} reduces to
\begin{equation}\label{ODEdim2}
\left\{
\begin{aligned}
\frac{dm}{d\tau} &= -2\, \frac{m + s}{s\,m - 1} m + 1 - m^2 - A \,, \\[4pt]
\frac{dh}{d\tau} &= \epsilon U( - m + s ) \,,\\[4pt]
\frac{dW}{d\tau} &=   \beta U \left(1 - m\, \frac{m + s}{s\,m - 1}\right)  - \epsilon\,(P_z/\rho+g)\,,\\[4pt]
\frac{ds}{d\tau} &= -\beta \frac{(1 + m^2)(1 + s^2)}{(1 - m s)}-\epsilon\frac{g + P_z (1 + m s)}{U}\,,\\[4pt]
\frac{d\beta}{d\tau} &= \frac{\beta (\beta +1) \left(1+m^2\right) s}{1 - m s} + \frac{1}{U}\frac{dC_e}{d\tau}-\epsilon (\beta+1) m P_z/U\,,\\[4pt]
\frac{dA}{d\tau} &= G_A\,,\\[4pt]
\frac{dC_{req}}{d\tau} &= G_{C_{req}}\,,\\[4pt]
\cdots\\
\cdots\\
\end{aligned}
\right.
\end{equation}
where $0<\epsilon\ll 1$ and the functions~$G_s,G_A,G_{C_{req}}$ are functions of other observables, such as velocity and pressure gradients. For example, from Eq.~\eqref{ODEdim}, their explicit expressions are given by
\begin{equation}\label{G_A}
\begin{aligned}
G_s& =-\beta \frac{(1 + m^2)(1 + s^2)}{(1 - m s)}-\epsilon\frac{g + P_z (1 + m s)}{U}\,,\\[12pt]
G_{A}&= \frac{4A \left(1+m^2\right) s}{1-m s}  + \frac{A}{U\beta} \left(\frac{dC_e}{d\tau} -\epsilon m P_z\right)+ A\epsilon^2 P_{xz} + \\
&  +\beta \epsilon^2\,U \left(2 m U_{xx} - U_{xz} (A -3 m^2 +1) - m U_{zz} (A  -m^2 +1 )\right) -\beta^2 \epsilon^2 \eta_{xxx}|_Q\, U^2\,,\\[12pt]
G_{C_{req}} &=\epsilon m P_z -\frac{(3+A) \beta \left(m^2+1\right) s U}{A (1-m s)} -\frac{\beta\, U P_{xz}}{A} + \\
&  + \frac{\beta^2\,U^2}{A^2}\left(-2 m  U_{xx} + m\,U_{zz} \left(A-m^2+1\right)+U_{xz} \left(A -3 m^2 +1\right)\right)  +\frac{\beta^3\, \eta_{xxx}|_Q \,U^3}{A^2}\,,\\[12pt]
G_{\beta}& =\frac{\beta (\beta +1) \left(1+m^2\right) s}{1 - m s} + \frac{1}{U}\frac{dC_e}{d\tau}-\epsilon (\beta+1) m P_z/U \,.
\end{aligned}
\end{equation}
Note that $G_A$ depends on the vertical pressure gradient, time derivative of the speed $C_e$, and third  order space derivatives of the wave surface. These depend on higher order gradients of velocity and pressure fields. 


A slow-fast timescale separation occurs near very steep focusing waves where $A\approx1$ and $C_e\approx C_{req}$. 
A detailed balance is attained if 
\begin{equation}\label{scalinga}
m,s\sim O(\epsilon^{1/2})\,,\quad A-1\sim O(\epsilon)\,,\quad C_e-C_{req}\sim O(\epsilon^{3/2})\,,\quad  h,P_z,U\sim~O(1)\,,
\end{equation}
and from the pressure equation~\eqref{eq:PoissonP} and pressure relation~\eqref{Pressurerelation2}, the second order pressure derivatives are scaled as 
\begin{equation}\label{scaling2a}
P_{xx},P_{zz}~\sim~O(\epsilon^{-2}),\qquad P_{xz}\sim~O(\epsilon^{-3/2})\,.
\end{equation}
Moreover, from Eq.~\eqref{Ce2}, near breaking the hotspot velocity~$C_e=C_{req}+O(\epsilon^{3/2})$ when $P_{xx}=P_{zz}+O(\epsilon^{3/2})$ and
\begin{equation}\label{scaling3a}
U_{xx},U_{zz},U_{xz},P_{xz}\sim~O(\epsilon^{-3/2})\,.
\end{equation}
Then, $dC_e/d\tau=dC_r/d\tau\sim O(\epsilon^{3/2})$. 

Using these scalings, the system in Eq.~\eqref{ODEdim2} simplifies as
\begin{equation}\label{ODEdim3}
\left\{
\begin{aligned}
&\it{Fast\,\,variables}\\[4 pt]
&\frac{dm}{d\tau} = \underbrace{2\, m s + m^2 + 1 - A}_{O(\epsilon)} + O(\epsilon^{3/2})\,, \\[6pt]
&\frac{ds}{d\tau} = \underbrace{\beta -\frac{g+P_z/\rho}{U}}_{O(\epsilon)} + O(\epsilon^{3/2})\,, \\[4pt]
&\frac{d(m+s)}{d\tau} = \frac{dm_2}{d\tau}=\underbrace{\left(1 - A +\beta -\frac{g+P_z/\rho}{U U_z}+m^2+2 m s\right)}_{O(\epsilon)} + O(\epsilon^{3/2})\,, \\[12pt]
&\it{Slow\,\, variables}\\[4 pt]
&\frac{dA}{d\tau} =\frac{d\beta}{d\tau}=  O(\epsilon^{3/2})\,,\\[4pt]
\end{aligned}
\right.
\end{equation}
where the leading terms on the right-hand sides of the fast equations are $O(\epsilon)$ and those of the slow equations are $O(\epsilon^{3/2})$. 
The slow-fast system is not closed as it depends on the pressure gradient $P_z$. This can be resolved from the observation that near breaking the momentum parameter $R= m +s\sim m_2$  is an adiabatic slow variable ~(see Fig.~\ref{FIG2_fastslow}). To make $R$ a slow variable we choose the pressure gradient 
\begin{equation}
    P_z/\rho= U U_z \left(1- A+\beta+m^2+2m s\right)-g\,,
\end{equation}
which is of $O(1)$. The vertical Lagrangian acceleration at the tracking point~$Q$ follows as
\begin{equation}\label{verta_L}
\begin{aligned}
    a_L &=(\partial_t W + U \partial_x W + W \partial_z W)\big |_Q\\[4pt]
    &=-(P_z/\rho + g ) = U U_z \left(A-1 - \beta-m^2-2m s\right)\,.
\end{aligned}
\end{equation}

The system simplifies to
\begin{equation}\label{ODEdim4}
\left\{
\begin{aligned}
&\it{Fast\,\,variables}\\[4pt]
&\frac{dm}{d\tau} = 2\, m s + m^2 + 1 - A + O(\epsilon^{3/2})\,, \\[4pt]
&\frac{ds}{d\tau} = -\frac{dm}{d\tau} + O(\epsilon^{3/2})\,, \\[12pt]
&\it{Slow\,\, variables}\\[4pt]
&\frac{dA}{d\tau} = \frac{dR}{d\tau}=\frac{d\beta}{d\tau}=O(\epsilon^{3/2})\,,\\[4pt]
\end{aligned}
\right.
\end{equation}

Thus, $m$ and $s$ are fast variables of $O(\epsilon^{1/2})$, $s=-m + O(\epsilon)$ and $A-1,\beta$ and $R\sim O(\epsilon)$ are adiabatic slow variables. Then, the evolution equation of the surface slope~$m$~(fast variable) is given by
\begin{equation}\label{foldeq}
   \frac{dm}{d\tau} = F(m,A)=- m^2 + 1 - A\,,  
\end{equation}
which is valid within the limits of $m\sim O(\epsilon^{1/2})$ and $A-1\sim O(\epsilon)$ with $A$ an adiabatic slow parameter. 

We conjecture that as the wave approaches breaking, its effective dynamics slow down, analogous to nearing a black-hole horizon, suggesting a gravity-analogue interpretation~\citep{Faccio2016Emergent}. In this view, higher-order pressure-gradient terms follow from imposing next-order slowness in the hierarchical perturbation expansion, without solving the nonlocal pressure equation~\eqref{eq:PoissonP}.

\section{Equivalent crest speed}\label{appC_eqcrestspeed}
Consider a generic moving point $Q$ on the surface~($\mathbf{X}_Q=(x_Q(t),z=\eta(x_Q(t),t)$) whose speed is $C=dx_Q/dt$ and the slope at $Q$ is $m(t)$. Then we can formulate the speed $C$ as that where
\[
\eta_x|_Q = m(t)\,. 
\]
The time derivative~\citep{FedeleJFM2016}
\[
\frac{d}{dt} (\eta_x(x_Q(t),t) -m(t)) =-\dot{m}  + \eta_{xx} C  + \eta_{xt} =0\,,
\]
vanishes at the moving point $Q$ and its speed is written as
\begin{equation}
    C =-\frac{\eta_{xt}|_Q - \dot{m}}{\eta_{xx}|_Q}\,. 
\end{equation}
The mixed derivative is given in Eq.~\eqref{etaxt} and
\begin{equation}\label{C0}
    C =U -\frac{U_z}{\eta_{xx}|_Q} +\frac{ 2\, U_x m +U_z m^2 - \dot{m}}{\eta_{xx}|_Q}\,, 
\end{equation}
where we have set $\eta_x=m$ in Eq.~\eqref{etaxt} because $Q$ follows the crest of the wave, and $U,U_x,U_z$ are the values at $Q$ of the horizontal fluid velocity and its gradients.

We now define the equivalent crest speed $C_{req}$ at $Q$ as if there were a crest at that point, i.e.,~$m=\dot{m} =0$. From Eq.~\eqref{C} it follows that
\begin{equation}\label{Cr}
    C_{req} = U -\frac{U_z}{\eta_{xx}|_Q} \,,
\end{equation}
and $C$ is written as
\begin{equation}\label{C}
   C =C_{req} +\frac{ 2\, U_x m +U_z m^2 - \dot{m}}{\eta_{xx}|_Q}\,.
\end{equation}
Near a focusing wave, where $\eta_{xx}|_Q<0$, $C_{req} \ge C$, with the equality attained at the wave crest.


  \bibliographystyle{elsarticle-harv} 
  \bibliography{jfm}



\end{document}